\documentclass[11pt,a4paper]{article}
\pdfoutput=1
\usepackage{jheppub}

\usepackage{epsfig,latexsym,amsfonts,amsmath,amsthm,amssymb,amsbsy,multirow,slashed,color,mathrsfs,wasysym,textcomp,subfigure,wrapfig,comment}

\def\del{\partial}
\def\nn{\nonumber}

\usepackage{color,datetime,graphicx}

\newcommand{\pink}[1]{\textcolor{\pink}{#1}}

\newcommand{\cale}{\mathcal{E}}

\title{Metastability in Bubbling AdS Space}

\preprint{LMU-ASC 44/14, IPhT-T14/103}

\author{Stefano Massai$^{a}$,}
\author{Giulio Pasini$^{b}$ and}
\author{Andrea Puhm$^{c}$}

\affiliation[a]{Arnold Sommerfeld Center for Theoretical Physics, Theresienstr. 37, 80333 Muenchen, Germany
}

\affiliation[b]{Institut de Physique Th\'eorique, CEA Saclay, 91191 Gif sur Yvette, France}

\affiliation[c]{Department of Physics, UCSB, Santa Barbara, CA 93106}

 \emailAdd{stefano.massai@lmu.de, giulio.pasini@cea.fr, puhma@physics.ucsb.edu}

 \abstract{
 We study the dynamics of probe M5 branes with dissolved M2 charge in bubbling geometries with
$SO(4) \times SO(4)$ symmetry. These solutions were constructed by Bena-Warner and Lin-Lunin-Maldacena and correspond to the vacua of the maximally supersymmetric mass-deformed M2 brane theory. 
We find that supersymmetric probe M2 branes polarize into M5 brane shells whose backreaction creates an additional bubble in the geometry. We explicitly check that the supersymmetric polarization potential agrees with the one found within the Polchinski-Strassler approximation.
The main result of this paper is that probe M2 branes whose orientation is opposite to the background flux can polarize into 
metastable M5 brane shells. These decay to a supersymmetric configuration via brane-flux annihilation.
Our findings suggest the existence of metastable states in the mass-deformed M2 brane theory.
}

\begin{document}
\maketitle

\section{Introduction}

String backgrounds holographically dual to non-conformal theories 
with reduced supersymmetry display very interesting infrared 
dynamics that give insight into string
theory resolution of supergravity
singularities and into strong coupling
and non-perturbative aspects of gauge
theories~\cite{Johnson:1999qt,Polchinski:2000uf,Klebanov:2000hb,
Maldacena:2000yy,Bigazzi:2003ui}. A celebrated example is the dual of the mass-deformed $\mathcal{N}=4$ SYM theory 
considered by Polchinski and
Strassler in~\cite{Polchinski:2000uf}, which involves polarization of D3
branes into shells of D5 or NS5 branes.
An important question is how to extend
these constructions to non-supersymmetric setups. One particularly
interesting problem is to find gravity solutions dual to metastable 
supersymmetry breaking.
This mechanism usually involves strong coupling dynamics in the gauge theory
and thus, in the absence of field theory dualities which permit to study the potential
at strong coupling (as in~\cite{Intriligator:2006dd}), the gravity description provides a most useful tool.

In this paper we will study metastable configurations in the gravity dual
of supersymmetric mass-deformed theories.
We will consider the M-theory analog of the still unknown Polchinski-Strassler
supergravity solution: the gravity dual of the maximally supersymmetric 
mass-deformed M2 brane theory constructed in~\cite{Bena:2004jw}, 
which involves polarization of M2 branes into M5 brane shells.
As we will review shortly, these solutions are part of a bigger class
of bubbling geometries in supergravity constructed
in~\cite{Lin:2004nb}, and they also share some similarities with the bubbling geometries constructed in~\cite{Bena:2005va,Berglund:2005vb,Bena:2007kg}. 
We anticipate that our findings
will provide evidence for metastable states in the
mass-deformed M2 brane theory, and are relevant more broadly for 
the construction of
non-supersymmetric bubbling geometries.

Our starting point is the supersymmetric solution that arises from giving equal masses to the four
hypermultiplets of the M2 brane theory. In the gravity picture this 
corresponds to a four-form
flux perturbation of $AdS_4\times S^7$, which causes an infrared
singularity~\cite{Pope:2003jp}. The resolution by polarization~\cite{Myers:1999ps} of the M2 branes into M5 brane shells has been studied by Bena
in~\cite{Bena:2000zb} within the Polchinski-Strassler approximation.\footnote{The field theory interpretation of the polarization in terms of a fuzzy three-sphere in ABJM theory~\cite{Aharony:2008ug} was given
in~\cite{Gomis:2008vc}.} Unlike the type IIB case, a solution which is
conjectured to capture the backreaction of
the M5 brane shells, and thus provide all the vacua of the mass-deformed
theory, has been constructed by Bena and Warner (BW)
in~\cite{Bena:2004jw}. A full understanding of the geometry of these
solutions was then given in terms of bubbling AdS space by Lin, Lunin
and Maldacena (LLM)~\cite{Lin:2004nb}. 
As we will review in detail, the
polarized M5 brane solutions are found by U-dualizing the LLM $1/2$ BPS
geometries in type IIB, which describe $\mathcal{N}=4$ SYM on $R \times S^3$ and
have an $R\times SO(4)\times SO(4)$ bosonic symmetry and 16
supersymmetries. The resulting eleven-dimensional solutions are
completely smooth and all the brane charge is dissolved in flux.
Different solutions corresponding to the fully backreacted concentric M5 brane shells are determined by the partition of the real line into black and white strips.

Our strategy will be to probe such geometries with M5 branes
made of polarized M2 branes with positive or negative charge, wrapping contractible
three-spheres inside various four-spheres of the geometry. 
When the probe M2 charge has the same orientation as the background flux, we find that the M5 brane potential has global supersymmetric
minima. These correspond to the classical supersymmetric vacua of the mass-deformed
theory, and we will see that they are geometrized precisely by an LLM
solution with an additional pair of black and white strips. This is a rather explicit
confirmation that the BW and LLM geometries describe
the backreaction of the dielectric M2 branes found
in~\cite{Bena:2000zb}. This also provides a nice check of the
Polchinski-Strassler magic: the result of probing the solution corresponding to 
fully backreacted M5 brane shells with dissolved M2 charge is precisely the 
same as probing the background that has all the M2 branes at the origin.

Allowing the probe M5 branes to carry M2 charge opposite to the charge
dissolved in the background fluxes of an LLM solution, we find that the M5 brane potential has
metastable minima close to the North Pole of one of the four-spheres
(near one of the strip boundaries of the LLM solution), for
some regime of parameters. Such a configuration decays via
non-perturbative bubble nucleation toward one of the supersymmetric
minima whose geometrization corresponds to another LLM solution (which has an additional pair of black and white strips). 
Our findings are reminiscent of the metastable
states in the Klebanov-Strassler and CGLP
backgrounds~\cite{Kachru:2002gs,Argurio:2006ny,Argurio:2007qk,Klebanov:2010qs} and in the bubbling backgrounds that correspond to black hole microstate geometries~\cite{Bena:2011fc}. 
The geometrization of our
metastable branes would give a non-supersymmetric bubbling geometry
dual to a metastable state in the mass-deformed M2 brane 
theory. In the type IIB frame, these configurations become
non-supersymmetric giant gravitons in $AdS_5\times S^5$. It would be very interesting to investigate the existence of metastable states in the field theory or by using some 
matrix model techniques. 

Our result is similar to the metastable supertubes found in~\cite{Bena:2011fc,Bena:2012zi} used to construct near-extremal black hole microstates, and suggests that the existence of metastable probe brane configurations in bubbling geometries is a generic feature.
Indeed, the bubbling backgrounds corresponding to supersymmetric black hole microstate
geometries~\cite{Bena:2005va,Berglund:2005vb,Bena:2007kg} 
share with the LLM solutions the feature that they contain backreacted polarized
branes. However, the dielectric probes we use here differ from the probe supertubes used in~\cite{Bena:2011fc,Bena:2012zi}, which correspond to M5 branes wrapping a contractible circle and carrying worldvolume flux coming from two orthogonal M2 branes and intrinsic angular momentum. Hence it is rather encouraging that the two probe calculations yield a very similar result. 
The backreaction of our metastable M5 branes with dissolved M2 charge, which we will comment on shortly, could then give insight into the even more challenging backreaction of the metastable supertubes. The corresponding supergravity solutions would provide evidence for the existence of a large class of non-supersymmetric black hole microstate geometries in the context of the fuzzball proposal
(see~\cite{Mathur:2005zp, Bena:2007kg, Mathur:2008nj, Balasubramanian:2008da, Skenderis:2008qn, Chowdhury:2010ct} and more recently~\cite{Bena:2013dka} for reviews).

In this paper we treat the M5 branes as probes, namely
we do not take into account their backreaction on the geometry. While
for the BPS probes we can easily identify the corresponding solution to be an
LLM geometry with an additional pair of black and white strip, for non-BPS probes the
backreaction is much more challenging. 
We mention that recently there has been much progress in the
construction of supergravity solutions for anti-branes in flux
compactification (see for
instance~\cite{McGuirk:2009xx,Bena:2009xk,Bena:2011wh,Massai:2011vi,Bena:2012bk,Giecold:2013pza}). In
the investigated cases it was shown that the corresponding
solution has infrared singularities whose resolution is currently a subject of intense study\footnote{For an account of the ongoing debate see for example~\cite{Blaback:2011nz,Bena:2012tx,Blaback:2012nf,Bena:2013hr,Bena:2012vz,Bena:2014bxa}.}. On the other hand, the UV asymptotics of such solutions are
in nice agreement with the interpretation as gravity duals of metastable
states~\cite{DeWolfe:2008zy,Bena:2011wh,Dymarsky:2013tna}. Our situation is dissimilar in various important aspects from the
geometries studied in the above mentioned references. One key point is
that our probes branes respect the symmetries of the background
geometries, which are themselves backreacted M5 branes with M2 charge
dissolved in flux. 
For this reason, we believe that in the present context the problem of
constructing the backreaction of the polarized non-BPS branes could be 
tractable. 
It would clearly be interesting to construct this challenging cohomogeneity-two solution 
which would give important insight into the physics of non-supersymmetric bubbling geometries.\\

The paper is organized as follows. In \S~\ref{sec:M} we discuss in detail the
M-theory bubbling solutions and we compute their M2 and M5 charges.
These solutions which are described by a sequence of black and white strips can be obtained by U-dualizing the  type IIB 1/2 BPS solutions
of LLM which we review in Appendix~\ref{appsec:bubblinggeo}.
In \S~\ref{sec:polarization} we show the IIA
reduction of the M-theory background and we use it to derive the Hamiltonian for a probe M5
brane with M2 charge dissolved in its worldvolume. We derive a limit of the Hamiltonian
which reduces the study of its minima to a one-dimensional problem.
In \S~\ref{sec:susymin} we study supersymmetric global
minima of the probe Hamiltonian. In particular we show that exact results we derive are in agreement with those obtained in the Polchinski-Strassler approximation in~\cite{Bena:2000zb}. 
We also show explicitly
that supersymmetric minima corresponding to polarized probes are geometrized by an LLM
solution with additional pairs of white and black strips at the location of the minima. 
In \S~\ref{sec:metastates} we show that the probe Hamiltonian admits
metastable configurations and we obtain an analytic expression for the position 
of the minima using a Polchinski-Strassler~--~type approximation. We then discuss 
the decay process of these metastable probes to supersymmetric minima which are in 
correspondence with the classical supersymmetric vacua of the mass-deformed M2 brane theory.
We illustrate the discussion of supersymmetric and metastable minima by plotting
the Hamiltonian for a particular example in \S~\ref{sec:susymin} and \S~\ref{sec:metastates} respectively.
We end with a discussion and a list of open problems in \S~\ref{sec:discussion}.
In the appendices we work out in detail how to obtain the M-theory solutions from
the type IIB LLM geometries, we give the relation between LLM and
BW solutions, correcting various typos in the literature,
and we give some more technical details.

\section{Gravity dual of the mass-deformed M2 brane theory}\label{sec:M}

In this section we discuss the solution of eleven-dimensional
supergravity dual to a mass-deformation of the
three-dimensional $\mathcal{N}=8$ worldvolume
theory of a stack of M2 branes. In general one can turn on masses for the
four hypermultiplets of the $\mathcal{N}=8$ theory, preserving four
supercharges. Note that contrary to the type IIB case, one cannot
add additional mass terms which explicitly break all the
supersymmetries. When all the masses are equal supersymmetry gets
enhanced and the mass-deformed theory preserves 16 supercharges. In the
dual picture, the mass deformation corresponds to perturbing $AdS_4
\times S^7$ by a non-normalizable four-form flux. This is the M-theory
analogue of the \mbox{$\mathcal{N}=1^{\star}$} perturbation of $AdS_5\times
S^5$ considered in~\cite{Girardello:1999bd}. In the latter case the
corresponding supergravity solution develops an infrared singularity
which 
was argued in the work of Polchinski and
Strassler~\cite{Polchinski:2000uf} to be resolved by brane
polarization~\cite{Myers:1999ps}. In the M-theory setup, one can
similarly show that the M2 branes that source $AdS_4 \times S^7$ can
be polarized by the four-form flux perturbation into shells of
M5 branes at finite $AdS$ radius and
extending in various planes inside the seven-sphere~\cite{Bena:2000zb}. The full
supergravity solutions corresponding to these polarized M5 brane
shells for the equal-mass perturbation have been obtained by Bena and Warner in~\cite{Bena:2004jw}. These solutions coincide with the U-duals of the type
IIB bubbling geometries constructed by Lin, Lunin and Maldacena
in~\cite{Lin:2004nb}, which permit to easily select boundary
conditions that lead to smooth regular solutions. 
In Appendix~\ref{appsec:bubblinggeo} we review
the original type IIB background and in Appendix~\ref{app:BW}
we report the explicit coordinate change between the BW and LLM solutions.
In this section we review this family of
regular supergravity backgrounds. In the next section we will
see explicitly that they correspond to the backreaction of
M5 brane shells with M2 charge dissolved in flux which are dual to the
vacua of the mass-deformed theory.

\subsection{M-theory bubbling geometries}\label{ssec:Mtheoryuplift}

The supergravity solution dual to the mass-deformed M2 brane theory is given by:\footnote{Note that the four-form field
 strength as given in (2.35) of~\cite{Lin:2004nb} is incorrect.}
\begin{align}
ds_{11}^2 &= H^{-2/3} (-dt^2 + d\omega_1^2+d\omega_2^2) +H^{1/3} \Big[
h^2 (dy^2 + dx^2) + y e^{G} d\Omega_3^2 + y e^{-G}
d\tilde \Omega_3^2\Big]\, , \label{11metric}\\
G_4&= -d(H^{-1} h^{-2} V) \wedge dt \wedge d\omega_1 \wedge
d\omega_2 \nn \\
&\quad +  \left[ d(y^2 e^{2G} V) - y^3 \star_2 dA \right]
\wedge d\Omega_3+  \left[d(y^2 e^{-2G} V) - y^3 \star_2
  d\tilde A\right] \wedge d\tilde \Omega_3\,. \label{11G4}
\end{align}
with warp factor $H$ given by
\begin{equation}
H=e^{-2\Phi} = h^2-V^2 h^{-2}\, .\label{IIAH}
\end{equation}
The metric describes a three-dimensional external space corresponding to the 
M2 brane worldvolume directions warped on an eight-dimensional transverse 
manifold that consists of a two-dimensional subspace spanned by the 
coordinates $(y,x)$ and two three-spheres $S^3$ and $\tilde{S}^3$. The Hodge star $\star_2$ refers to the flat space spanned by $(y,x)$.
The functions $A, \tilde{A},h,G,V$ are given by
\begin{align}
&A = \frac{z+\frac{1}{2}}{y^2}\, , \quad
\tilde{A}= \frac{z-\frac{1}{2}}{y^2}\, , \\
&h^{-2} = 2 y \cosh G \, , \quad
G = {\rm arctanh}(2z)\,,\label{Gtoz}\\
&y \partial_y V =\partial_x z\, , \quad
y \partial_x V = - \partial_y z\, . \label{eq:V} 
\end{align}
The full solution is determined in terms of a single master function $z(x,y)$ that obeys a linear equation:
\begin{equation}
 \partial_x^2 z + y \partial_y \left( \frac{\partial_y z}{y}
 \right) =0\, . \label{eq:z}
\end{equation}
The coordinate $y$ plays a special role since it is the product of the radii of the
two three-spheres. At $y=0$ at least one of the two three-spheres shrinks to zero size. For the geometry to be smooth, the shrinking three-sphere and the radial direction should combine to form $\mathbb{R}^4$. This requires the function $z$ to have a special behavior. The geometry is non-singular if the boundary values of equation~\eqref{eq:z} are $z = \pm \frac{1}{2} $ on the $y=0$ line spanned by $x$. As long as $y \neq 0$ there are two non-vanishing three-spheres, $S^3$ and $\tilde S^3$. At the $y=0$ line, $S^3$ shrinks to zero in a non-singular fashion if $z=-\frac{1}{2}$, while $\tilde S^3$ shrinks smoothly if $z=\frac{1}{2}$. Both spheres shrink at the boundary of these two regions where they combine to form $\mathbb{R}^8$.

\begin{figure}[ht!]
\centering
\subfigure[]{
 \includegraphics[width=0.3\textwidth]{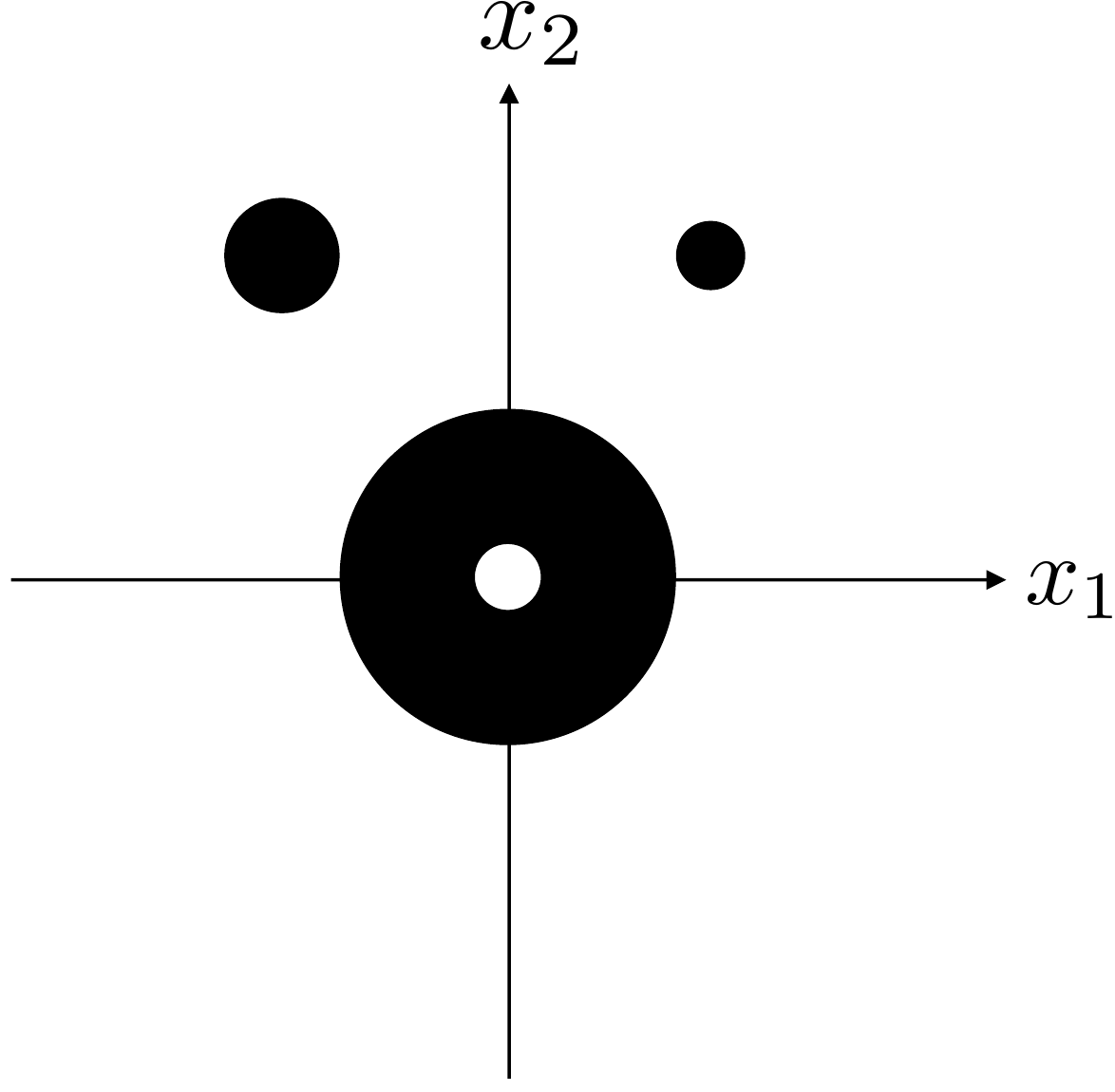}
 }
 \hspace{2cm}
 \subfigure[]{
\includegraphics[width=0.2\textwidth]{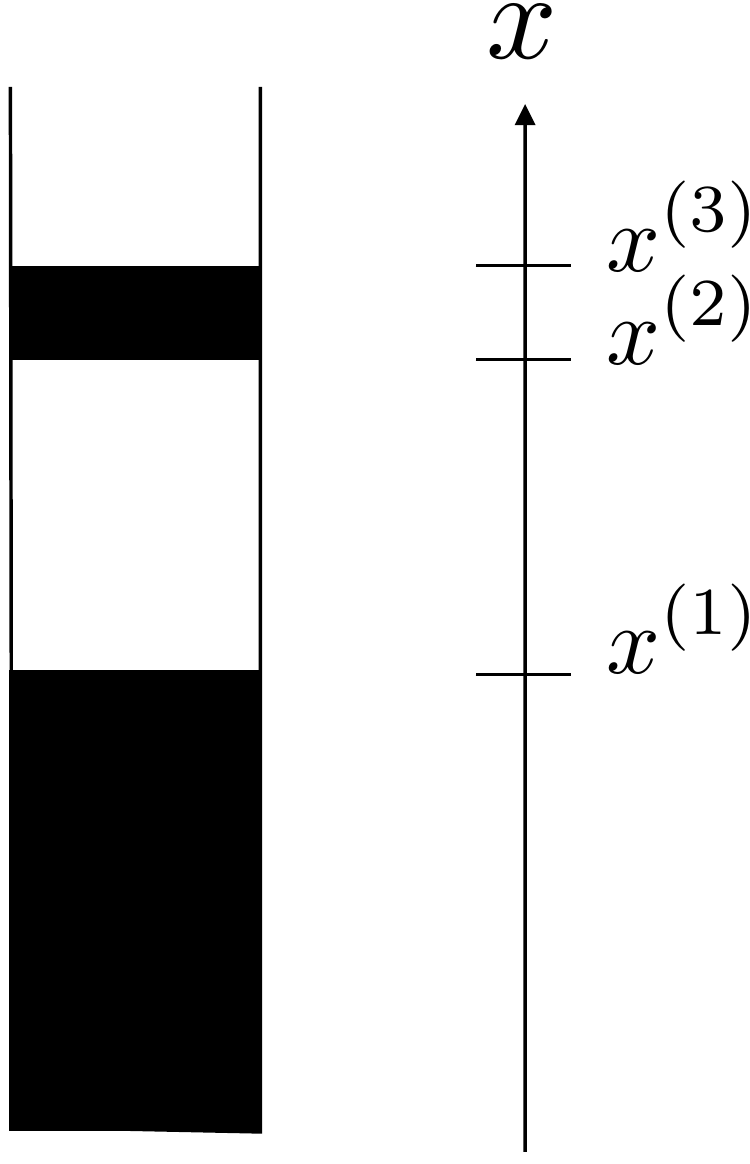}
 }
\caption{(a) A general type IIB solution is defined by boundary
   values of $z$ in the $y=0$ plane spanned by $(x_1,x_2)$, depicted as black and
 white droplets. (b) A general solution of the mass-deformed M2 brane theory, 
 obtained as a  superposition of plane waves, which are limits of the type IIB droplets.}
\label{fig:blackwhite}
\end{figure}

One way to pictorially represent the boundary behavior of $z$ in the $y=0$
line is by drawing black and white strips according to the value of
$z=\pm \frac{1}{2}$. We depict this black and white partitioning of
the real line $x$ in Figure~\ref{fig:blackwhite} (b). Such
configurations can be obtained from the U-dualization of the type IIB
LLM solutions (see Appendix \ref{appsec:bubblinggeo}), which are in correspondence with a black and white coloring
of a two-plane
spanned by coordinates $(x_1,x_2)$ as shown in
Figure~\ref{fig:blackwhite} (a). 

 \subsection{The multi-strip solution}\label{ssec:multistrip}

A general smooth solution is determined by a superposition of solutions to \eqref{eq:V} and
\eqref{eq:z} with the boundary value of $z$ being $\pm 1/2$:
\begin{align}
 z_0(x,y) &= \frac{1}{2} \frac{x}{\sqrt{x^2+y^2}}\,,\label{zstrip} \\
 V_0(x,y) &= -\frac{1}{2} \frac{1}{\sqrt{x^2+y^2}}\,.
\end{align}
In the type IIB frame this solution corresponds to the half-filled plane and the metric is that of a plane wave.
For the metric \eqref{11metric} to asymptote to $AdS_4\times S^7$, the
multi-strip solution must have a semi-infinite black region at one
side of the $y=0$ line and a semi-infinite white region on the
other. The simplest non-trivial solution corresponds to a pair of
finite-size white and black strips with adjacent semi-infinite black
and white regions, represented in Figure~\ref{fig:blackwhite} (b).
A general multi-strip solution is then obtained by superposition:
\begin{align}
z(x,y)&=\sum_{i=1}^{2s+1} (-1)^{i+1} z_0(x-x^{(i)},y)\,, \label{multiz}\\
V(x,y)&= \sum_{i=1}^{2s+1} (-1)^{i+1} V_0(x-x^{(i)},y)\,,\label{multiV}
\end{align}
where $x^{(i)}$ is the position of the $i$th boundary and $s$ denotes the number of pairs of white and black strips. For odd $i$ the
boundary changes from black to white while for even $i$ the boundary
changes from white to black. This will be the
general form of a smooth solution corresponding to classical supersymmetric of
the mass-deformed theory.

\subsection{M2 and M5 charges}\label{ssec:M2M5charges}

\begin{figure}[ht!]
\centering
\subfigure[]{
 \includegraphics[width=0.4\textwidth]{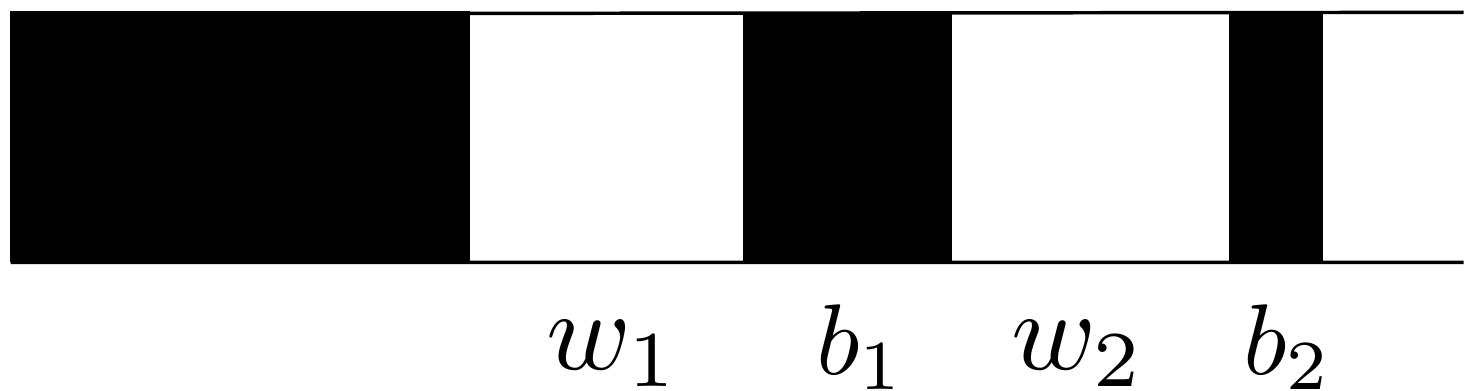}
 }
 \hspace{2cm}
 \subfigure[]{
\includegraphics[width=0.2\textwidth]{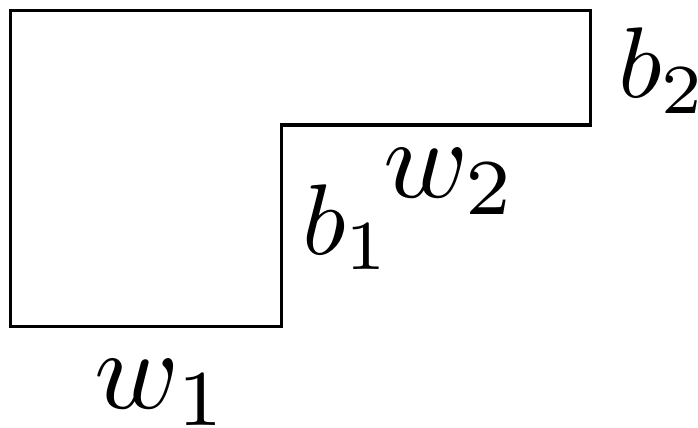}
 }
\caption{Correspondance between the partition of the real line that
  defines a general solution (a) and the Young diagram (b),
  illustrated for a two-strips solution.}
\label{fig:young}
\end{figure}

We now show that the metric~\eqref{11metric} indeed asymptotes to $AdS_4 \times S^7$.
With the coordinate transformation
\begin{equation}
y =\frac{R^2}{2}\sin \alpha \, , \qquad
 x = \frac{R^2}{2}\cos \alpha\,, \label{coordinatechange1}
\end{equation}
the two three-spheres combine with the angle $\alpha$ to form a
seven-sphere. For large radii $R$ the warp factor $H$ reduces to the
warp factor of an M2 brane and we recover the standard harmonic M2
brane metric:
\begin{equation}
 ds^2=H^{-2/3}dx_{\parallel}^2+H^{1/3} dx_{\perp}^2\, , \qquad H
 =\frac{32\pi^2 l_p^6 N}{R^6}  \, ,
 \end{equation}
where $l_p$ is the Planck length in eleven dimensions and $N$ denotes
the M2 charge as given by 
\begin{equation}\label{M2Maxwell}
 N=\frac{1}{(2\pi l_p)^6}\int_{S^7_\infty} \star_{11} G_4\, ,
\end{equation}
where $S^7_\infty$ is a seven-sphere in the asymptotic region. 
From the expansion of the warp factor of the multi-strip solution introduced in the
previous section we get:
\begin{equation}
R^6 H = 8\sum_{i=1}^{s} \Big[ (x^{(2i+1)}-x^{(2i)}) \sum_{j=1}^{i}
(x^{(2j)}-x^{(2j-1)})\Big] \equiv 8 T \, ,
\end{equation}
from which we get that the M2 charge of the solution is related to the
strip widths as:
\begin{equation}\label{NfromUV}
 N = \frac{T}{4\pi^2 l_p^6} \, .
\end{equation}
As discussed in~\cite{Lin:2004nb}, a useful way to represent an LLM
geometry is through the Young diagram corresponding to the momentum
basis of free fermions. 
It is easy to see that $T$ corresponds to the number of boxes of the
Young diagram associated to the particular configuration. 
For a general multi-strip solution, the black and white regions map to the
vertical/horizontal edges of the Young diagram with the edge sizes
corresponding to the sizes of the respective strips (see
Figure~\ref{fig:young} for an illustration). 

\begin{figure}[h!]
\centering
\includegraphics[width=0.3\textwidth]{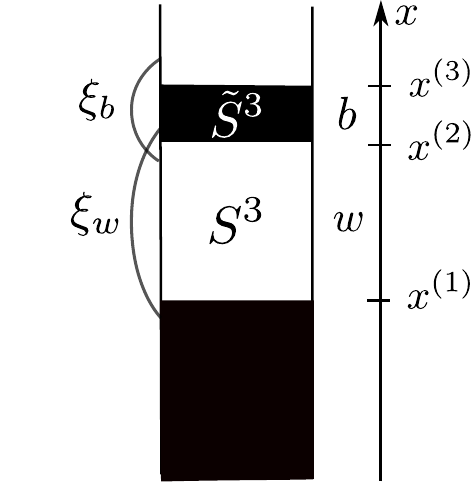}
\caption{The M5 charge corresponding to a white (black) strip is obtained by integrating the four-form flux over a four-cycle obtained by fibering the $S^3$ ($\tilde S^3$) over the curve $\xi_w$ ($\xi_b$) whose end points lie in a region where the $S^3$ ($\tilde S^3$) shrinks to zero size.}
\label{fig:stripS3arc}
\end{figure}

We stress that $G_4$ does not tend to the standard harmonic solution
({\it i.e.} with legs along the M2 worldvolume only) as it also contains
two additional transverse terms. These are the non-normalizable modes
associated to the mass perturbation in the dual M2 brane theory. These transverse
fluxes give rise to an M5 dipole charge:
\begin{equation}
M=\frac{1}{(2\pi l_p)^3}\int_{S^4} G_4\,.\label{M5charge}
\end{equation}
There are various topological 4-cycles in the multi-strip solution. 
For example, we can consider an
$S^4$ containing an $S^3$, which is obtained by fibering the $S^3$ on
a curve $\xi_w$ that encloses a white strip and whose boundary ends at $y=0$ on a region where $z=-1/2$,
{\it i.e.} where the $S^3$ smoothly shrinks to zero size, as
illustrated in Figure~\ref{fig:stripS3arc}.

 For definiteness, let us
consider the first white strip of length $w=x^{(2)}-x^{(1)}$. 
We then obtain, from~\eqref{11G4}:
\begin{equation}
(2\pi l_p)^3 M_w =  2\pi^2 \int_{\xi_w}\Big[ d(y^2 e^{2G} V) - y^3
  \star_2 dA \Big]\, .\label{eq:QM5}
\end{equation}
The function $y^2 e^{2G} V$ is smooth and globally well-defined and so the first term in \eqref{eq:QM5} does not contribute to the integral via Stokes theorem. 
The second term in \eqref{eq:QM5} satisfies the Laplace equation
$d(y^3 \star_2 dA)=0$ and we obtain:
\begin{equation}\label{eq:Q_M5}
(2\pi l_p)^3 M_w =  4\pi^2 \int_{x^{(1)}}^{x^{(2)}} dx
\left(z+\frac{1}{2} \right)\Big|_{z=1/2}= 4\pi^2
\left(x^{(2)}-x^{(1)}\right) = 4\pi^2 w\,.
\end{equation}
We learn that the M5 charge corresponding to the four-form flux
through the $S^4$ is proportional the size of the white
strip $w$. Similarly, the M5 charge $M_b$, corresponding to four-form flux
through an $\tilde S^4$ containing an $\tilde S^3$, obtained by
fibering the $\tilde S^3$ on a curve $\xi_b$, is proportional to the
size of the black strip $b$:
\begin{equation}\label{M5b}
(2\pi l_p)^3 M_b = 4\pi^2 b \, .
\end{equation} 
Clearly, the same result also applies for a general multi-strip
solution, which contains various 4-cycles. From charge quantization, this result also gives the
quantization condition for the length of the strips and it agrees with
the result found in~\cite{Cheon:2011gv}. 

As a further check of our normalizations, we note that we can compute the M2 charge
of the solution~\eqref{M2Maxwell} from the IR data, using the transverse fluxes. We can
do this by deforming the $S^7$ to the IR region $y\approx 0$,  according to $S^7_\infty=  \mathcal{D}_7+\partial \mathcal{M}_8$, where $\mathcal{D}_7$ is a shrinking region with $y \approx 0$ and $\mathcal{M}_8$ is spanned by $(y,x)$ and the two three-spheres. Since the geometry is smooth the integral over $\mathcal{D}_7$ vanishes and \eqref{M2Maxwell} reduces to
\begin{equation}\label{M2charge}
(2\pi l_p)^6 N=  -\frac{1}{2}  \int_{\mathcal{M}_8}G_4 \wedge G_4\, ,
\end{equation}
where we used the equation of motion $d\star_{11} G_4 = -\frac{1}{2}
G_4 \wedge G_4$. The integral on the right hand side
of~\eqref{M2charge} can be shown to factorize into products of M5
charges over the various four-spheres of a general multi-strip
solution, given in~\eqref{eq:Q_M5} and~\eqref{M5b}. By taking into
account the correct orientation of the fluxes, there are cancellations that lead
precisely to the result~\eqref{NfromUV}, expressing $N$ in terms of
the number of blocks of the corresponding Young diagram.
We note that for the solution corresponding to a single pair of
finite-size black and white strips of length respectively $w$ and $b$,
the M2 charge is simply given by 
\begin{equation}\label{M2chargeonestrip}
N= \frac{w b}{4\pi^2 l_p^6}\, , 
\end{equation}
which indeed corresponds to the number of boxes of a rectangular Young
diagram after taking into account the normalization~\eqref{eq:Q_M5}.

\section{Probe M5 brane Hamiltonian}\label{sec:polarization}

In this section we derive the Hamiltonian for a probe M5 brane
with M2 charge dissolved in its worldvolume whose dynamics will be discussed
in \S~\ref{sec:susymin} and \S~\ref{sec:metastates}.
In particular, we will classify its global and local minima as a function of its charges. In \S~\ref{sec:susymin} we will
discuss supersymmetric minima, which provide a nice check that the
background solution is indeed sourced by dielectric branes. 
In \S~\ref{sec:metastates} we will show that when the probe M2 charge is opposite 
to the M2 charge of the background the probe M5 brane has non-supersymmetric 
metastable minima in some regime of parameters. 
Our discussion will be general and valid for an LLM solution
with an arbitrary number of strips. We will focus on 
a regime in which we can trust simple approximations of the probe brane
potential which will suffice to prove the existence of global and
local minima. We will also give numerical results for a simple example
which will help to clarify  the general discussion. 

There are two ways to study the M5 polarization potential, either directly
via the Pasti-Sorokin-Tonin action~\cite{Pasti:1997gx} (as done for example in~\cite{Bena:2000zb})
or via the Dirac-Born-Infeld -- Wess-Zumino action for D4 branes probing the 
type IIA reduction of the M-theory solution.
Both yield the same result and we will use the second one since it is more clear.

\subsection{type IIA reduction}\label{ssec:MIIA}

To compute the potential for M2 branes polarizing into M5 branes it is convenient to work with the type IIA reduction of the M-theory solution
\eqref{11metric}-\eqref{11G4} along $\omega_2$. We relegate a detailed discussion of the type IIA
solution to Appendix~\ref{appssec:IIA} and summarize here the result. The
metric and fluxes are:
\begin{align}
ds^2_{IIA}&= H^{-1} (-dt^2 + d\omega_1^2) + h^2(dy^2 + dx^2) +  y e^{G
} d\Omega_3^2 + y e^{ - G} d \tilde \Omega_3^2\, ,\label{MIIAmetric}\\
B_2 &= -H^{-1} h^{-2} V dt \wedge d\omega_1\, ,\\
F_4 &=  \left[d(y^2 e^{2G} V) - y^3 \star_2 dA\right]
\wedge d\Omega_3 +  \left[d(y^2 e^{-2G} V) - y^3 \star_2
  d\tilde A\right] \wedge d\tilde \Omega_3\label{MIIAF4} \, .
\end{align}
To compute the polarization potential in the next section we will also
need the explicit expressions for
the RR gauge potentials $C_3$ and $C_5$. 
Since $C_1=0$ we have $F_4=dC_3$ and $\star F_4 =  F_6 =  dC_5 + H_3 \wedge C_3$. 
In the multi-strip solution \eqref{multiz}-\eqref{multiV} we can solve these equations analytically.
For $C_3$ with legs on the $S^3$ we have
\begin{equation} \label{MIIAC3}
c_3(x,y)= \sum_{i=1}^{2s+1} (-1)^{i+1}
 \frac{2 (x-x^{(i)})^2+y^2}{2\sqrt{(x-x^{(i)})^2+y^2}}+x+ y^2 e^{2G(x,y)} V(x,y) +c\, .
\end{equation}
In the next section we will discuss in detail the role of the
constant $c$ which corresponds to a gauge choice for $c_3$.
The RR five-form potential $C_5$ for the multi-strip solution with legs on the $S^3$ is
\begin{equation}\label{MIIAC5}
 c_5(x,y)=\frac{2y^2}{1-2 z(x,y)}- y^2+ \frac{c_3(x,y) V(x,y)}{H(x,y) h(x,y)^{2}}\, .
\end{equation}
Similar expressions are obtained for the RR forms with legs along $\tilde S^3$ (see Appendix~\ref{appssec:IIA}).

\subsection{The probe action}\label{ssec:action}

We are interested in the potential for M5 branes carrying M2 charge in the M-theory solution discussed in \S~\ref{sec:M}. The same potential is obtained from probe D4 branes carrying F1 charge placed in the dimensionally reduced IIA solution discussed in \S~\ref{ssec:MIIA}.
Hence we consider a D4 brane wrapped on a three-sphere of the internal space
and which carries dissolved F1 charge along $\omega_1$. 
The embedding is given by $t=\sigma^0$, $\omega_1=\sigma^1$ and
$\sigma^2$, $\sigma^3$, $\sigma^4$ along the three-sphere.
The probe D4 brane action 
is given by
\begin{align}
S_{D4} &= - \mu_4 \int d^5 \sigma e^{-\Phi} \Big[ -\det
\left(g_{ab}+2\pi\alpha'\mathcal{F}_{ab}+B_{ab}
\right)\Big]^{1/2} \nn \\ 
&\quad + 
\mu_4 \int \Big[ C_5 +(2\pi\alpha' \mathcal{F}_2+B_2)\wedge
C_3\Big] \, ,
\end{align}
where $\mathcal{F}_2$ is the induced worldvolume field strength on the brane
\begin{equation}
\mathcal{F}_2=\mathcal{E} d\sigma^0 \wedge d\sigma^1\, ,
\end{equation}
and $\mu_4$ is the D4 brane tension
\begin{equation}\label{mu4}
\mu_4 = \frac{2\pi}{g_s(2\pi l_s)^5} = \frac{1}{(2\pi)^3 \mu_{1}l_p^3}\, ,
\end{equation}
which, for future use, is expressed in terms of the F1
string tension $\mu_{1} = 2\pi\alpha'$ and the eleventh dimensional
Planck length $l_p$.
In the background~\eqref{MIIAmetric} with RR gauge potentials given
by~\eqref{MIIAC3} and~\eqref{MIIAC5} we obtain after integrating on
the three-sphere $S^3$:
\begin{equation}
S_{D4} = \int d^2\sigma \mathcal{L}(\mathcal{E}) \, ,
\end{equation}
with
\begin{equation}
\mathcal{L}(\mathcal{E}) = -\mu_4 V_{S^3} \Big[ y^{3/2} e^{3G/2} H^{1/2}
\sqrt{H^{-2}-(\cale+{B}_2)} + \, c_5 + (\cale + {B}_2)
  c_3\big]
\end{equation}
where $V_{S^3}$ is the volume of the three-sphere spanned by $\sigma^2$, $\sigma^3$ and $\sigma^4$ and we recall that the warp factor $H$ is given by $H=h^2-V^2
h^{-2}$.
In order to compute the potential for the D4 brane we need to express
the Lagrangian in terms of the F1 charge, which is proportional to the
electric displacement~\cite{Herzog:2002yu,Klebanov:2010qs}:
\begin{equation}\label{ndef}
n =\frac{\del \mathcal{L(\mathcal{E})}}{\del
  \cale}\equiv \mu_{1}V_{S^3}\mu_4 \, p\, .
\end{equation}
The Hamiltonian is obtained from the Legendre transformation:
\begin{equation}
\mathcal{H} = n \cale - \mathcal{L(\mathcal{E})}\, .
\end{equation}
This gives the potential for D4 branes with dissolved F1 charge or,
equivalently, the potential for M5 branes with dissolved M2 charge:
\begin{equation}\label{eq:Hamiltonian}
\mathcal{H}= \mu_4 V_{S^3} \Big[ H^{-1} \sqrt{H y^3 e^{3G} + \left(p - c_3\right)^2} -p{B}_2 - c_5\Big]\,.
\end{equation}
In the subsequent sections we will study the dynamics of M2 branes polarizing into M5 brane
probes as described by this Hamiltonian. Note that we can also consider polarization into 
multiple M5 branes. The Hamiltonian for $m$ M5
branes is obtained multiplying~\eqref{eq:Hamiltonian} by an overall
factor $m$ and replacing $p \to p/m$.

While we will focus on M5 branes wrapping the $S^3$, a similar analysis
can be carried out for M5 branes wrapping the $\tilde S^3$. To obtain
the Hamiltonian one just has to replace $G\to -G$ and $V_{S^3} \to
V_{\tilde S^3}$ in~\eqref{eq:Hamiltonian} and substitute $\tilde c_3$
and $\tilde c_5$ for the RR fields whose expression is given in
Appendix~\ref{appssec:IIA}.

To avoid cumbersome notation coming from the normalization~\eqref{ndef} and~\eqref{eq:Hamiltonian}, we will simply drop the overall factor in ~\eqref{eq:Hamiltonian} and in the rest of the paper, when the distinction between $n$ and $p$ is not crucial, we will refer to $p$ as the M2 charge.

\subsection{One-dimensional Hamiltonian} \label{ssec:Hyto0limit}

\begin{figure}[ht!]
\centering
\includegraphics[width=0.6\textwidth]{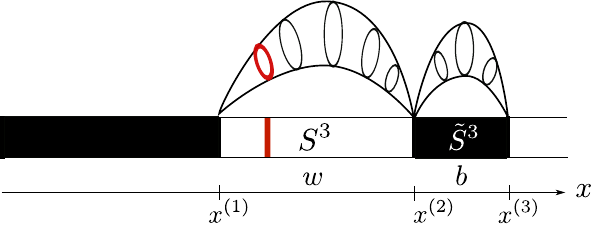}
\caption{
The topology of a single pair of finite-size white and black strips that are smoothly connected to a semi-infinite black strip on the left boundary and to a semi-infinite white strip on the right boundary. We consider (in red) probe M5 branes with dissolved M2 branes wrapping the $S^3$ that remains of finite size in the white strip region.}
\label{fig:topology}
\end{figure}

To study the minima of the potential \eqref{eq:Hamiltonian} of a probe M5 brane
wrapping the $S^3$ of a multi-strip solution we substitute $c_3$ and $c_5$
with \eqref{MIIAC3}-\eqref{MIIAC5}. 
It can be shown that the Hamiltonian minimizes on
the $y=0$ axis, when either one or both of the three-spheres shrink to
zero size. We can thus reduce the problem to finding the explicit form of the Hamiltonian
in one dimension, on the $y=0$ line. Since we are considering an M5 brane wrapping the background
$S^3$, the interesting dynamics will happen inside white strips where $S^3$ is of finite size. We will thus focus on the $y \to 0$ limit of the Hamiltonian in the region of the real line where the master function $z$ takes the value $+1/2$. 
When approaching a white strip, the function $z$ behaves as
\begin{equation}
z(x,y) = \frac12 - y^2 \zeta_+^2(x) + \mathcal{O}(y^4) \, ,
\end{equation}
which defines the function $\zeta_+^2(x)$. For a multi-strip solution (see
\S~\ref{ssec:multistrip}), this function is given by
\begin{equation}
\zeta_+(x) = \frac12
\sqrt{\sum_{i=1}^{2s+1}(-1)^{i+1}\frac{|x-x^{(i)}|}{(x-x^{(i)})^3}}\, .
\end{equation}
The function $V(x,y)$ then approaches $V_+(x)$:
\begin{equation}
V_+(x) = -\frac{1}{2}\sum_{i=1}^{2s+1} \frac{(-1)^{i+1}}{|x-x^{(i)}|}  \, .
\end{equation}
The warp factor and the B-field
can be expressed as follows:
\begin{equation}\label{Hplus}
H_+(x) = \frac{\zeta_+^2(x) -V_+^2(x)}{\zeta_+(x)} \, ,\qquad B_+(x) =
-\frac{V_+(x)}{\zeta_+^2(x)-V_+^2(x)} \, ,
\end{equation}
The three-form gauge potential approaches
\begin{equation}
c_3^+(x) =  \sum_{i=1}^{2s+1} (-1)^{i+1}|x-x^{(i)}|+ 
x + \frac{V_+(x)}{\zeta_+(x)^2} + c \, .\label{eq:c3+}
\end{equation}
where the integration constant $c$ corresponds to a gauge choice and the five-form gauge potential approaches
\begin{equation}
c_5^+(x) = \frac{1}{\zeta_+^{2}(x)} - c_3^+(x) B_+(x)\,.
\end{equation}
We give the details of this derivation in Appendix~\ref{appsec:limits}. 
The Hamiltonian for a probe M5 brane wrapping the $S^3$ restricted to white strips on the $y=0$ line is then given by 
\begin{equation}\label{hamplus}
\mathcal{H}_+ (x)= H_+(x)^{-1} \sqrt{\frac{H_+(x)}{\zeta_+^3(x)} +
  \left[p-c_3^+(x)\right]^2} - B_+(x)\Big[p-c_3^+(x)\Big] -\frac{1}{\zeta_+^{2}(x)} \, .
\end{equation}
In the following we study the global and local minima of this Hamiltonian for the multi-strip solution of \S~\ref{ssec:multistrip}. We are most interested in probe M5 branes carrying M2 charge that polarize inside white strips at finite distance from the strip boundaries. This is illustrated for the simplest bubbling solution in Figure~\ref{fig:topology}.

One can also consider the $y\to 0$ limit of the Hamiltonian in the black region where the $S^3$ wrapped by the probe M5 brane shrinks to zero size. Naively one would expect the potential to vanish inside this region since the M5 brane has shrunk to zero size. Due to the non-trivial structure of supersymmetric M2 brane minima, which we will discuss in \S~\ref{ssec:degsusy}, this is not the case in general and we will study what happens inside black strips in \S~\ref{ssec:Dirac}.

As already pointed out in \S~\ref{ssec:action} we can also consider the potential for probe M5 branes wrapping the $\tilde S^3$. The one-dimensional Hamiltonian for this case for both the black and the white regions can be found in Appendix \ref{appsec:limits}.

\section{Supersymmetric minima: DBI meets SUGRA}\label{sec:susymin}

We now look for supersymmetric minima of the probe
Hamiltonian~\eqref{hamplus} that describes M5 branes wrapping the $S^3$ and is restricted to the white strip regions of the real line $y=0$.
To satisfy $\mathcal{H}_+=0$ we have to impose
\begin{equation}\label{susymin} 
 \left|c_3^+(x) - \frac{V_+(x)}{\zeta_+(x)^2}-p \right| \zeta_+(x)-\left(c_3^+(x) - \frac{V_+(x)}{\zeta_+(x)^2}-p\right)V_+(x)=0\, . 
 \end{equation}
As we will show, there are two different ways to solve~\eqref{susymin}. Correspondingly, there exist two different kinds of minima: those where the probe M5 brane
shrinks to an M2 brane, and those
where the M5 retains a finite-size. This second class of minima proves that the building blocks
of our background are indeed M5 branes with dissolved M2 branes and are the analogue of the ones found in~\cite{Bena:2000zb}. 

\subsection{Degenerate minima}\label{ssec:degsusy}

To satisfy~\eqref{susymin} we observe that
\begin{equation}
\lim_{x \to x^{(i)}} \frac{V_+(x)}{\zeta_+(x)}= (-1)^i \, ,
\end{equation}
which means that the probe Hamiltonian can
have supersymmetric minima located at the boundaries
$x^{(i)}$ of the strips. This can easily be understood as follows. At the strip boundaries
both $S^3$ and $\tilde{S}^3$ shrink to zero size, and our probe M5 brane reduces to an M2 brane. As the background is maximally
supersymmetric and sourced by dielectric M2 branes, a probe M2 feels zero force if it preserves all the 16 supercharges, i.e. if it has the same orientation as the dielectric M2 branes of the background. To fully solve~\eqref{susymin} for $x=x^{(i)}$ we notice that $\frac{V_+}{\zeta_+^2}=0$ at the boundaries. Defining the effective M2 charge
\begin{equation}\label{peff}
p^{eff}_+(x^{(i)})=p-c_3^{+}(x^{(i)})\,,
\end{equation}
 we see that the Hamiltonian has a supersymmetric minimum at the
boundary $x^{(i)}$ if
\begin{align}
p^{eff}_+(x^{(i)})&>0 \quad \textmd{($i$ odd)} \, ,\qquad \qquad p^{eff}_+(x^{(i)})<0  \quad
\textmd{(i even)}\, . \label{degeneratemin}
\end{align}
The physical meaning of $p^{eff}_+$ is clear: inserting the M5 probe
in a white strip close to a boundary $x^{(i)}$, part of its M2 charge $p$ is screened
by the value of the potential $c_3^{+}(x^{(i)})$. Indeed, from~\eqref{hamplus}
we see that the effective M2 charge of the probe close to a boundary
is $p^{eff}_+$ rather than $p$.

Eq.~\eqref{susymin} shows that M2 branes are BPS at odd boundaries, while anti-M2 branes are BPS at even boundaries. Another way to check this is to plot
the potential for M2/anti-M2 probes which, using $G_4=dA_3$, is given by 
\begin{equation}\label{M2ham}
\mathcal{H}_{M2/anti-M2}=  H^{-1} \mp A_{012}=(h^2 \mp V)^{-1}\,.
\end{equation}
This potential has indeed minima at the $y=0$ line at odd
or even boundaries respectively for $-$ or $+$ in~\eqref{M2ham}. This is also confirmed by the analysis of the supersymmetry projector~\cite{Cheon:2011gv,Hashimoto:2011nn}.

\subsection{Polarized minima}\label{ssec:polsusy}

The second way to solve~\eqref{susymin} is to require the expression inside the absolute value and the brackets to vanish. This yields for the location of the supersymmetric minima:
\begin{equation}\label{susy_minimum_exact}
x_{susy}=\frac{1}{2} \left(p+x^{(1)}+
\Sigma_b^l - \Sigma_b^r-c\right)\, ,
\end{equation}
where $\Sigma_b^l$ and $\Sigma_b^r$ are the total size of the black
strips that are respectively to the left and right of the white strip
in which the probe M5 brane polarizes. 
In addition to the degenerate minima, we see from~\eqref{susy_minimum_exact} that the Hamiltonian has minima located at a finite distance away
from the boundaries. In the following, we will explicitly prove that
these are the minima that become, upon backreaction, LLM bubbling solutions corresponding to the classical supersymmetric vacua of the mass-deformed M2 brane theory.

Depending on the value of the constant $c$ in~\eqref{susy_minimum_exact} such minima exist for positive as well as negative induced M2 charge $p$. The value of this constant corresponds to the gauge choice used to describe the physics at the supersymmetric minimum. We will come back to this gauge choice in detail in \S~\ref{ssec:decay} where we need to understand the effect on the probe brane when changing gauge. In the remainder of this section
we will fix the gauge suitably to avoid cumbersome notation.

We mention that a result similar to~\eqref{susy_minimum_exact} applies
as well for M5 branes wrapping the $\tilde S^3$ which is non-vanishing inside
black strips. There are thus two channels into which a collection of
(anti-) M2 branes can polarize: either into an M5 brane wrapping the $S^3$ or into an
M5 brane wrapping the $\tilde S^3$. These are the different
polarization channels that arise in the probe
analysis~\cite{Polchinski:2000uf,Bena:2000zb}. Since the analysis for the two channels is 
analogous, from now on we will focus on polarization inside the white strips. Polarization
inside the black strips will be important in
\S~\ref{ssec:decay} to describe the final supersymmetric
configuration metastable branes can decay to.

As a final remark, we stress that~\eqref{susy_minimum_exact} holds also for the Hamiltonian for $m$ M5 branes, provided that one replaces $p$ with $p/m$.

\subsubsection*{Expansion \`{a} la Polchinski-Strassler}

We now want to show that one can get the same result~\eqref{susy_minimum_exact} by expanding the Hamiltonian at large distances from the branes sourcing the background. 
As discussed in \S~\ref{sec:M}, in this region the M-theory solution approaches
the $AdS_4\times S^7$ background perturbed by the four-form fluxes transverse to the M2 brane worldvolume directions, corresponding to the mass deformation in the dual M2 brane theory. 
The minimum we will find momentarily by expanding the Hamiltonian in the geometry of backreacted M5 branes is in agreement with the minimum found in~\cite{Bena:2000zb} where the four-form fluxes were treated as perturbation of $AdS_4 \times S^7$.
Note that in~\cite{Bena:2000zb} the M5 brane potential
was investigated directly using the Pasti-Sorokin-Tonin
action~\cite{Pasti:1997gx}, while here we are recovering the same
result using the type IIA reduction.

Starting from the full Hamiltonian~\eqref{eq:Hamiltonian} it is
convenient to first perform the coordinate change~\eqref{coordinatechange1}, expand the Hamiltonian at large $R$ and then define
\begin{equation}
r^2=R^2 \cos\left(\frac{\alpha}{2}\right)\,, \qquad \tilde r^2=R^2 \sin\left(\frac{\alpha}{2}\right)\,,
\end{equation}
where $r$ is the radius of $S^3$ and $\tilde r$ is the radius of
$\tilde{S}^3$. These radii are related to the original $x$ and $y$
coordinates as
\begin{equation}
y = r \tilde r \, ,\qquad 2x = r^2-\tilde r^2 \, .
\end{equation}
In this way one can get an approximate expression for the
Hamiltonian in the ultraviolet. As the probe is wrapping the $S^3$ the Hamiltonian minimizes for $\tilde r=0$, {\it i.e.} for $\alpha \to 0$, which coincides with the $y \to 0$ limit of \S~\ref{ssec:Hyto0limit}. Hence for $r$ large and $\tilde r=0$ one has for the metric functions appearing in \eqref{eq:Hamiltonian}
\begin{equation}
 H^{-1} \sim \frac{r^6}{N}+r^2\,, \qquad H y^3 e^{3G} \sim N\,,
\end{equation}
and for the form fields
\begin{equation}
 B_2 \sim \frac{r^6}{N}+\frac{r^2}{2}\,, \qquad c_3 \sim \frac{2N}{r^2} \,, \qquad  c_5\sim - r^4\,,
\end{equation}
where $N$ is related to the M2 charge of the background given by \eqref{M2charge}. 
Inserting these expansions into \eqref{eq:Hamiltonian}, the Hamiltonian reduces to:
\begin{equation}\label{Hamiltonian_approximate1}
\mathcal{H}\sim \left(\frac{r^6}{N}+r^2\right)\sqrt{N+\left(p-\frac{2N}{r^2}\right)^2}-\frac{pr^6}{N}-\frac{pr^2}{2}+r^{4}
\end{equation}
In~\cite{Bena:2000zb} the probe is taken to have a much larger M2
charge than M5 charge.  This reduces to the requirement $p >> \sqrt{N}$,
which allows to Taylor expand the square root in \eqref{Hamiltonian_approximate1}
to get the final result\footnote{One can also get this result directly by expanding the
Hamiltonian~\eqref{hamplus} for large $x$ setting $2x \sim r^2$ and keeping only the leading terms in $1/p$.}
\begin{align}\label{Hamiltonian_approximate}
\mathcal{H} &\sim \frac{pr^2}{2} -r^4+ \frac{r^6}{2p}\nn\\
&= \frac{r^2}{2p}\left(r^2-p\right)^2\,,
\end{align}
which is in perfect agreement with the result
of~\cite{Bena:2000zb}. Notice that the two higher order terms $\sim p r^6/N$ in
\eqref{Hamiltonian_approximate1} representing the M2 brane potential
cancel out, and $\mathcal{H}$ is a perfect square as
expected because of supersymmetry. The
Hamiltonian~\eqref{Hamiltonian_approximate} has a minimum for $r^2 = p$,
which is nothing but~\eqref{susy_minimum_exact} in the
ultraviolet.

Restoring the correct mass dimension $\mu$ that comes with the
four-form flux perturbation, one can check that the $r^4$ term is
linear in $\mu$, while
the $r^2$ term has mass dimension $\mu^2$  (see for example \S 4.2
of~\cite{Bena:2014bxa} for a simple review of the holographic origin
of the polarization potential~\eqref{Hamiltonian_approximate}).
Note that this term cannot be
explicitly computed in the Polchinski-Strassler~--~type analysis performed
in~\cite{Bena:2000zb}, since in that case the background is
computed only to first order in the transverse flux perturbation and
thus only at linear order in $\mu$. However, it can be correctly guessed from supersymmetry just by completing the square
and our result confirms this rather explicitly.\footnote{In type IIB,
  the $AdS_5\times S^5$ background perturbed by three-form fluxes at
  second order has been computed in~\cite{Freedman:2000xb,Taylor:2001pp}, reproducing the PS result.}

We can actually say much more. 
In the previous discussion we focused
on the UV region and so we neglected the widths of the LLM strips in
the IR. However, our analysis is not restricted to the asymptotic
region. Firstly, the expression~\eqref{Hamiltonian_approximate} also approximates the
Hamiltonian for \emph{small} $x$, {\it i.e.} near a strip boundary $x^{(i)}$, if we
identify $2(x-x^{(i)})\sim r^2$ in~\eqref{hamplus}. 
The location of the minimum is then in agreement
with~\eqref{susy_minimum_exact}. The reason why the probe potential is described by
the same expression~\eqref{Hamiltonian_approximate} inside finite-size strips is easy to understand. The $r^6$ term comes from the three-sphere the probe M5 is wrapping, and so this term is the same for both types of white strips. 
For the $r^4$ term things are much less obvious
and naively this term seems to depend on the details of the
backgrounds. However, by the magic noticed in~\cite{Polchinski:2000uf,Bena:2000zb}, 
this term only depends on the UV
boundary conditions, since it comes from an expansion of a form field which
is both closed and co-closed. The $r^2$ term then is fixed by
supersymmetry and hence is again the same for both types of strips. This is indeed the
reason why one can safely compute the brane polarization by putting
all the M2 branes at the origin: when they are puffed-up the probe will still
feel the same potential. Again, since we are now probing the full
geometry we can check this rather explicitly.

\subsubsection*{DBI versus SUGRA}

The fact that our probe potential correctly reproduces the result
of~\cite{Bena:2000zb} is a strong check that the Bena-Warner and LLM
solution describe the backreaction of M5 branes
polarized by the transverse four-form fluxes. Indeed we can see that
the probe analysis is in full agreement with the supergravity
solution. Consider an arbitrary LLM solution with strips
located at boundaries $x^{(1)} , \dots ,x^{(2s+1)}$, and let us focus on
the asymptotic region very far from the strips, {\it i.e.} $x>>
x^{(2s+1)}$. The previous analysis shows that a
probe M5 brane with dipole charge $m$ and with large M2 charge $n$, will polarize in this region
at
\begin{equation}\label{locationM2}
x \approx \frac {n/m}{2\mu_{1}\mu_4 V_{S^3}} \, ,
\end{equation} 
where we wrote $p$ in terms of the probe charge by using~\eqref{ndef}.
What is the supergravity solution corresponding to this probe M5
brane? It is easy to show that this solution is found by adding an
additional black strip carrying M5 charge $M_b=m$ , precisely at the
location~\eqref{locationM2}. In fact, the M2 charge of such solutions
is, using the
relations~\eqref{M5b},~\eqref{M2chargeonestrip} and~\eqref{mu4}:
\begin{equation}
N \approx \frac {n/m}{2\mu_{1}\mu_4 V_{S^3}} \times \frac{M_b}{2\pi
  l_p^3}  = n \, ,
\end{equation}
which nicely matches the M2 charge of the probe.
Hence, this explicitly confirms that the LLM solutions indeed geometrize the
supersymmetric minima found in the probe limit.
A similar, though more involved, correspondence between DBI and SUGRA was studied for supertubes in bubbling backgrounds in~\cite{Bena:2008dw,Bena:2013gma}.

Repeating the same reasoning for the case of the
supersymmetric minima~\eqref{susy_minimum_exact} that arise inside the
white strips is straightforward but more tedious. The backreaction of probe branes located at those minima is again described by an LLM solution with an additional pair of white
and black strips.

 We stress that a completely similar analysis can be carried out for
 supersymmetric minima that arise for M5 brane probes wrapping the
 $\tilde S^3$ which is non-vanishing inside the black strips.

\subsubsection*{Example: Bubbling solution with a single pair of white
  and black strips}

We now specialize the previous discussion to a simple example.  We focus on the simplest LLM geometry containing dielectric branes, namely the
solution corresponding to a single pair of finite-size white and black strips and 
we consider the dynamics of probe M5 branes within the white
strip, {\it i.e.}  M5 branes wrapping the $S^3$ in the
M-theory solution~\eqref{11metric}-\eqref{11G4}.
The white region of interest is smoothly connected to a semi-infinite black strip 
on the left boundary and to a finite-size black strip on the right boundary which 
smoothly connects to a semi-infinite white strip. 
We denote by $w=x^{(2)}-x^{(1)}$ and $b=x^{(3)}-x^{(2)}$,
respectively, the widths of the finite-size white and black strip (see
Figure~\ref{fig:topology}). Without
loss of generality we set $x^{(1)}=0$ and we fix the gauge so that $c_3^+(0)=0$. 

We first discuss degenerate supersymmetric minima that arise at the
boundary of the strips. On the left boundary of the white strip ($x=0$) the Hamiltonian simplifies to
\begin{equation}
 \mathcal{H}_+(0) = \left(\left|p\right|-p\right) \frac{w(w+b)}{b}\, .
\end{equation} 
Hence for $p \geq 0$ the Hamiltonian has a supersymmetric minimum at the
left boundary, where the $S^3$ the M5 brane is wrapping shrinks to zero
size. On the right boundary of the white strip ($x=w$) the Hamiltonian simplifies to
\begin{equation}
 \mathcal{H}_+(w)=
 \left[\left|2w-p\right|-\left(2w-p\right)\right]
 \frac{w b}{(w+b)}\, ,
\end{equation}
and hence for $p\leq 2w$ the Hamiltonian has a supersymmetric
minimum at the right boundary. 
Note that $c_3^+(0)=0$ and $c_3^+(w)=2w$ and so we have $p^{eff}=p$ on
the left boundary and $p^{eff}=p-2w$ on the right boundary. Hence, the conditions 
on $p$ to have supersymmetric minima at the boundaries
are precisely the conditions that $p^{eff}>0$ on the left boundary and
$p^{eff}<0$ on right boundary as discussed in \S~\ref{ssec:degsusy}.

\begin{figure}[ht!]
\centering
\subfigure[$p=80$.]{
 \includegraphics[width=0.43\textwidth]{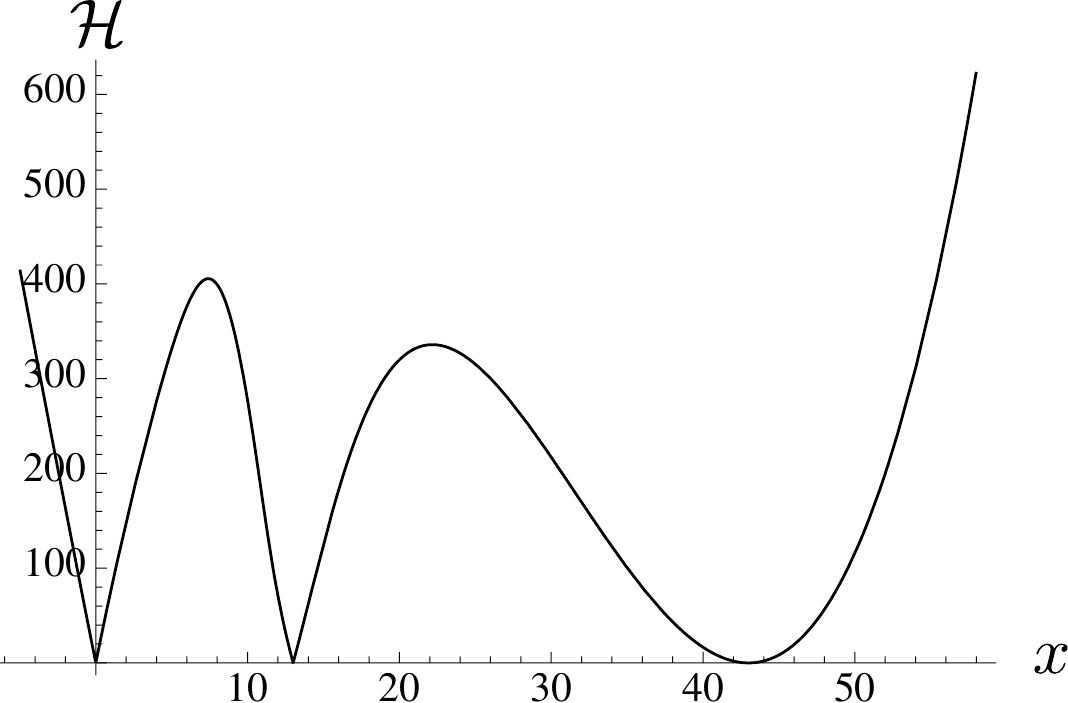}
 }
 \hspace{1cm}
\subfigure[$p=8$.]{
 \includegraphics[width=0.43\textwidth]{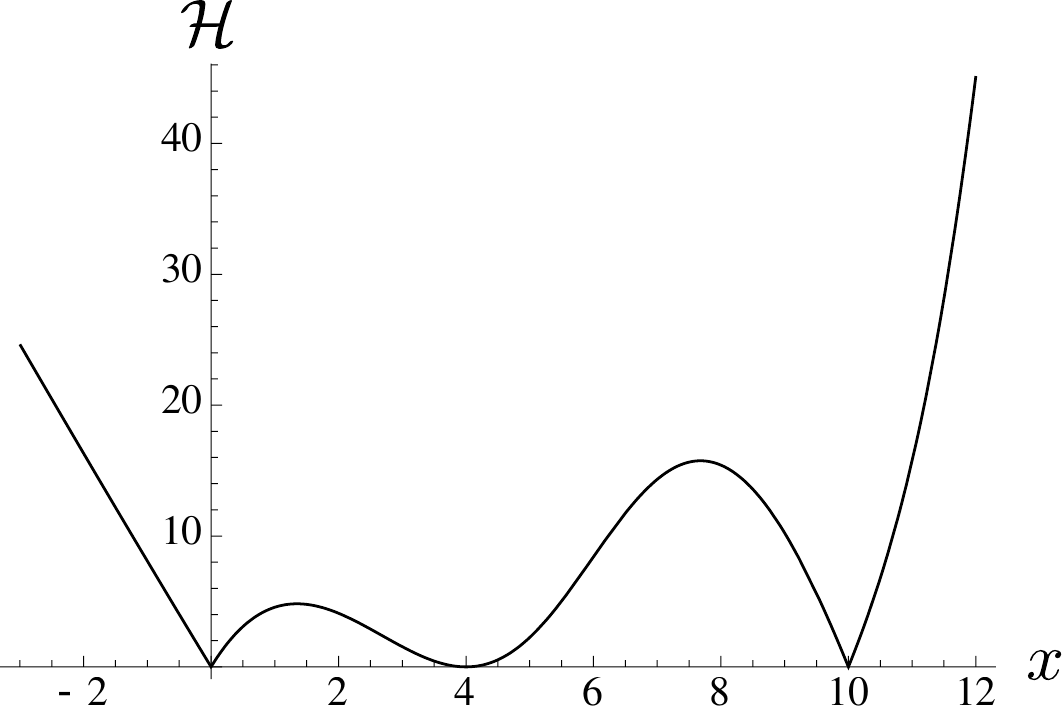}
 }
\caption{Supersymmetric global minima of the probe potential,
  illustrated for a solution with $w=10$ and $b=3$. (a) A supersymmetric
  minimum in the semi-infinite white strip; the minima in this
  asymptotic region correspond to those found
  in~\cite{Bena:2000zb}. (b) A supersymmetric minimum inside the white strip.}
\label{fig:susymin}
\end{figure}

We expect that probe
M2 branes placed at the boundaries of the white strip will polarize
into BPS M5 branes at a finite distance from the boundaries, as illustrated in Figure~\ref{fig:topology}. The
backreaction of these probe branes is captured by an LLM
geometry with an additional pair of black and white strips. The
general result~\eqref{susy_minimum_exact} for the position of such
supersymmetric minima now simplifies to:
\begin{equation}
x_{susy}=\left\{\begin{aligned}&\frac{p}{2}\,, \;\; \qquad \text{finite
     size white strip}\\ & b+ \frac{p}{2}\,, \quad
   \text{semi-infinite white strip} \end{aligned}\right.
\end{equation}
We show the minimum in the asymptotic region and the minimum inside
the white strip in Figure~\ref{fig:susymin}.

\subsection{Wrapped Dirac strings}\label{ssec:Dirac}

So far we have discussed the Hamiltonian for a probe M5 brane wrapping the $S^3$, which is of finite size inside white strips. The probe can stabilize at a finite distance from a boundary inside a white strip or has degenerate minima at the boundaries of the strip where $S^3$ shrinks to zero size. Inside black strips the probe reduces to an M2 brane and the Hamiltonian is thus determined by the dynamics of this M2 brane. 
In the following we explain what happens inside black strips.

In an analogous way as for white strips we can take the $y \to 0$ limit of the Hamiltonian \eqref{eq:Hamiltonian} for black strips, {\it i.e.} for regions where the master function $z$ takes the value $-1/2$. We refer to Appendix~\ref{appssec:limitblack} for details and state here the result:
\begin{equation}\label{hamminus}
 \mathcal{H}_-(x)=\frac{1}{\zeta_-(x)^2-V_-(x)^2}\left[ \zeta_-|p-c_3^-(x)|+V_-(x)(p-c_3^-(x))
 \right] \, ,
\end{equation}
with $V_-(x)=V_+(x)$ and $\zeta_-(x)$ given by~\eqref{app:zetaminus}.
The three-form potential reduces to
\begin{equation}\label{c3minus}
c_3^-(x) = \sum_{i=1}^{2s+1} (-1)^{1+i}|x-x^{(i)}|+ 
x +c= x^{(1)} + 2 \Sigma_w + \Sigma_b +c\,,
\end{equation}
where $s$ is the number of pairs of finite-size white and black strips of the configuration,
$\Sigma_w$ is the total width of white strips to the left of the
black strip in which we study the Hamiltonian and $\Sigma_b$ is the
total width of black strips in the solution. Note that the three-form potential is constant inside black strips.

The Hamiltonian \eqref{hamminus} is considerably simpler than the Hamiltonian \eqref{hamplus} because the M5 brane is of zero size inside black strips and, hence, the Hamiltonian is dictated by the dynamics of the M2 branes.
From \eqref{hamminus} we see that the Hamiltonian vanishes inside a black strip if the M2 charge of the probe equals the value of the three-form potential inside that black strip. We can understand this as follows.
The effective M2 charge $p^{eff}_-(x^{(i)})=p-c_3^-(x^{(i)})$ corresponds to the M2 charge at the boundary $x^{(i)}$ of a black strip. Hence, if $p^{eff}_-(x^{(i)})=0$ there are no M2 branes at the boundary $x^{(i)}$ and the Hamiltonian~\eqref{hamminus} describing ``nothing'' vanishes everywhere inside that black strip.

If the effective M2 charge inside the black strip is non-zero, 
the situation is more complicated.
Recall from the discussion of degenerate minima of the Hamiltonian in \S~\ref{ssec:degsusy} that the probe M2 brane potential~\eqref{M2ham} has minima at the $y=0$ line at odd or even strip boundaries depending on whether the effective M2 charge \eqref{peff} is positive or negative. 
Hence, for non-zero values of the M2 charge, the
Hamiltonian~\eqref{M2ham} vanishes only at one of the boundaries of
the black strip. The Hamiltonian inside the black strip is then determined by the potential felt by M2/anti-M2 branes:
\begin{equation}\label{flatpot}
 V_{M2/anti-M2}=|p^{eff}_-| \mathcal{H}_{M2/anti-M2}\, .
\end{equation}
One can indeed check that the Hamiltonian~\eqref{hamminus} coincides with the potential felt by M2 branes if $p^{eff}_->0$ while it coincides with the potential felt by anti-M2 branes if $p^{eff}_-<0$.

\begin{figure}[ht!]
\centering
\subfigure[$p=0$.]{
\includegraphics[width=0.4\textwidth]{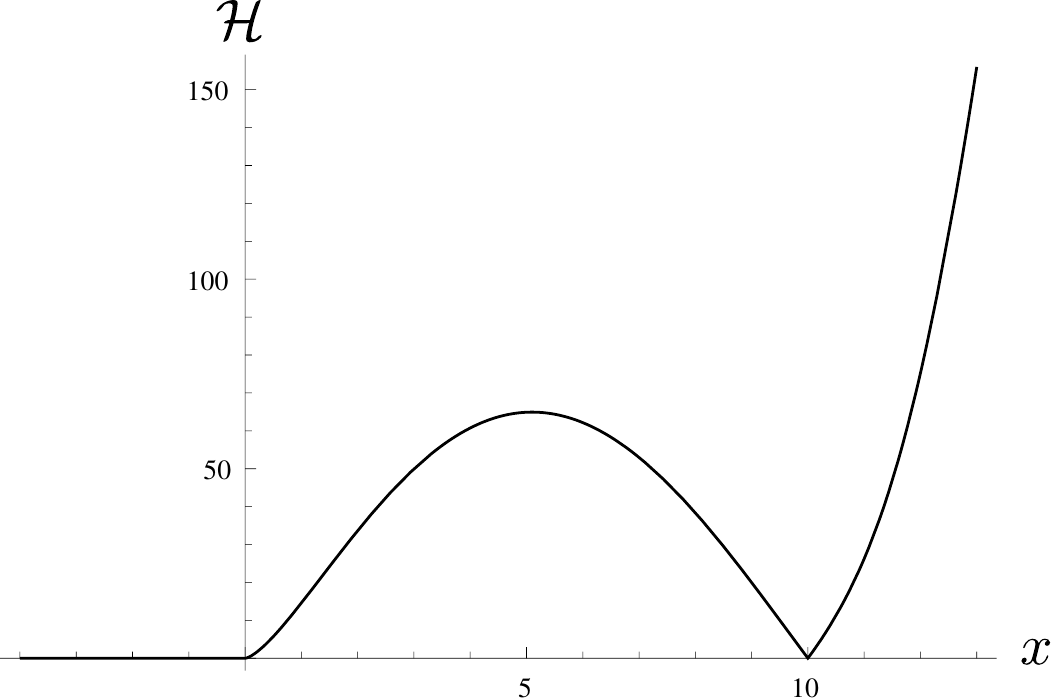}
}
\hspace{1.5cm}
\subfigure[$p=20$.]{
\includegraphics[width=0.4\textwidth]{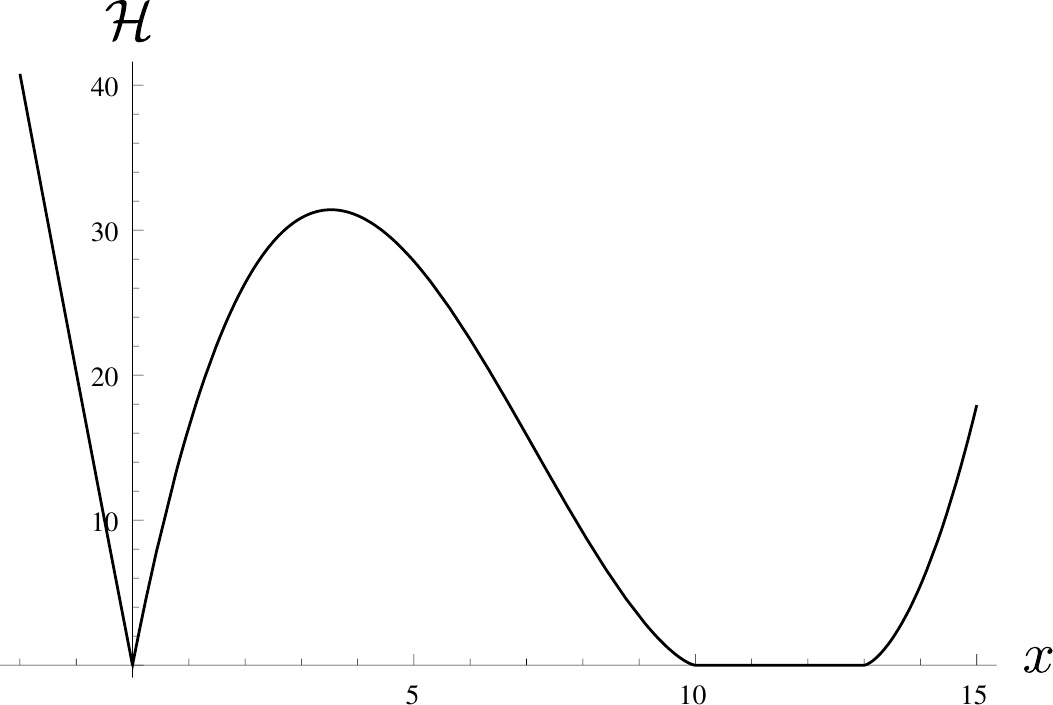}
}
\caption{The Hamiltonian in black strips describing ``nothing''.
The Hamiltonian vanishes inside the semi-infinite black strip for $p=0$ while it vanishes in the finite-size black strip for $p=2w$.}
\label{fig:flattening}
\end{figure}

We illustrate the flattening for the example of the single pair of white and black strip 
introduced in~\S~\ref{ssec:polsusy}. The semi-infinite black strip and the
finite-size black strip are located respectively at $-\infty<x<0$ and
$w < x<w+b$ on the $y=0$ axis (see Figure~\ref{fig:topology}) where the three-form potential~\eqref{c3minus}
takes the constant values $b+c$ and $2w+b+c$, respectively.
Choosing $c=-b$ yields a gauge where $c_3^+(0)=0$ and consequently $c_3^+(w)=2w$. 
The M5 brane Hamiltonian then vanishes inside the semi-infinite black strip for $p^{eff}(x^{(1)})=0$ which implies $p=0$. 
The Hamiltonian vanishes inside the finite-size black strip for $p^{eff}(x^{(2)})=0$ corresponding to $p=2w$.
We illustrate this for $w=10$ and $b=3$ in Figure \ref{fig:flattening}.

\section{Metastable M5 branes}\label{sec:metastates}

In this section we study local minima of the
Hamiltonian~\eqref{hamplus} that are not supersymmetric. We will focus
on the white strip $[x^{(2i-1)},
x^{(2i)}]$. As we will show, according to the value of $p$
in~\eqref{hamplus} there can be metastable minima close to the left
boundary $x^{(2i-1)}$ or close to the right boundary $x^{(2i)}$ of the
strip. In order to avoid clutter we will fix the gauge such that 
$c_3^+=0$ at the boundary of the strip we are expanding around which implies
$p^{eff}_+=p$ at that boundary.
For definiteness, we will focus on metastable minima close to $x^{(2i+1)}$
with $p$ negative, so that the probe is no longer BPS at the left boundary of a white strip. 
We first derive analytic expressions that approximate well the
location of such local minima, by using a Polchinski-Strassler~--~type
of expansion. We then focus on the simple example of a single pair of
white and black strip and we study the full Hamiltonian
numerically. We end with a discussion of the decay process for
metastable probes.

\subsection{Analytic results}\label{ssec:anametastates}

In order to get analytic control over the M5 brane Hamiltonian, we would
like to Taylor expand it around the boundary $x^{(i)}$, with $i$ odd and $p$ negative.
While this expansion can be rather cumbersome, we should realize that for
small enough $|p|$, many terms are actually subleading. Hence, it is sensible to keep only those terms
that are of the leading order in $p$ at the minimum. For
$x^{(i)}<x<x^{(i)} +|p|$ the Hamiltonian \eqref{hamplus} is well approximated by 
\begin{equation}\label{approximationp}
\mathcal{H}_+ \approx -p \left[ B_+(x) + \frac{1}{H_+(x)}\right] + c_3^+(x)\left[ B_+(x) + \frac{1}{H_+(x)}\right] -\frac{1}{\zeta_+^{2}(x)} - \frac{1}{p} \frac{1}{2\zeta_+^3(x)}
\, .
\end{equation}
This is nothing but the familiar form of the potential for polarized
branes. The linear in $p$ term is the force felt by probe anti-M2 branes in
the background geometry, the constant in $p$ piece comes from the 
$p$-independent Wess-Zumino action
and the inverse in $p$ term comes from the metric of the wrapped three-sphere. 
Starting from this expression, one can Taylor expand around
$x^{(i)}$, keeping in mind that it is enough to keep only the leading
terms. This can be easily achieved by noticing that 
\begin{equation}
-\frac{1}{2 \zeta_+^3(x)} = -4 (x-x^{(i)})^3 +
\mathcal{O}\left((x-x^{(i)})^5\right) \, ,
\end{equation}
and
\begin{equation}
- \left[B_+(x) + \frac{1}{H_+(x)}\right]  = 
 a_1 +  a_2 (x-x^{(i)}) + \mathcal{O}\left((x-x^{(i)})^2\right)  \,,
\end{equation}
where $a_1$ and $a_2$ are constants whose values depend on $x^{(i)}$: 
\begin{align}\label{coefficients}
  a_1 &=2\left(\sum_{j=1,j\neq i}^{2s+1}
   \frac{(-1)^{j}}{|x^{(i)}-x^{(j)}|}\right)^{-1}  \\
a_2 &= \frac{3}{4}
\left(\sum_{j=i+1}^{2s+1}\frac{(-1)^{j}}{(x^{(i)}-x^{(j)})^2}-\sum_{j=1}^{i-1}\frac{(-1)^{j}}{(x^{(i)}-x^{(j)})^2}\right)\left(a_1\right)^2
\, .
 \nn 
\end{align}
Writing $2(x-x^{(i)}) \approx r^2$ in terms of the
radius $r$ of the wrapped three-sphere $S^3$ we finally see that~\eqref{hamplus} is well-approximated for small $r$ and
small $|p|$ by:
\begin{equation}\label{Hpgeneral}
\mathcal{H_+} \approx p \,a_1 + p \, \frac{a_2}{2} \, r^2 - \frac{1}{2p}r^6 \, .
\end{equation}
If $a_2>0$ the Hamiltonian~\eqref{Hpgeneral} always has a metastable minimum at
\begin{equation}\label{minima}
r^2=|p| \sqrt{\frac{a_2}{3}} \, .
\end{equation}
We can explicitly check that the terms of the potential~\eqref{Hpgeneral} are
detailed balanced, namely at the minimum the last two terms scale with
the same power of $p$. One can also check that the omitted terms scale
at the minimum with sub-leading power of $|p|$. 

We would like to comment on an important difference between the
metastable probe potential~\eqref{Hpgeneral} and the supersymmetric
potential~\eqref{Hamiltonian_approximate}. In the latter case, the
minimum arises from a balance of $r^2$, $r^4$ and $r^6$ terms which
combine to give a perfect square. In the present case, the $r^4$ term
of the potential is missing, and the polarization is caused by the
\emph{negative} $r^2$ term. This term comes from the imperfect cancelation
of gravitational attraction and electric repulsion that the anti-M2
probes feel in the background. In our case the term is negative since
anti-M2s are repelled from the left boundary $x^{(i)}$, thus making the
polarization more likely. This is clearly very different from the
usual supersymmetric Polchinski-Strassler~--~type of dynamics, where the
polarization is caused just by a negative $r^4$ term, coming from the
Wess-Zumino action alone.

Recently (see~\cite{Bena:2014bxa}), a negative $r^2$ term has also been found in the 
potential for anti-M2 branes polarizing into M5 branes at the tip of a
warped Stenzel space~\cite{Stenzel:1993,Cvetic:2000db}. This analysis takes into account the full
backreaction of the anti-M2 branes on the geometry, 
and hence a repulsive force on probe anti-M2 branes is a signal of a tachyonic instability. 
We remark that in the present situation we work in a probe approximation, 
and thus we cannot easily draw conclusions regarding the negative $r^2$ term
felt by a probe anti-M2 brane.
It would be extremely interesting to investigate
the fate of our polarization potential once the full backreaction of
the probe on the LLM geometry is taken into account. 
We will come back to this point in \S~\ref{sec:discussion}.

When $|p|$ grows, the approximation~\eqref{Hpgeneral} breaks down and we would need to
keep next-to-leading order pieces in order to study the behavior of the
potential. While this can be done, the general result is rather
cumbersome, so we will postpone the discussion to a particular example in
the next section. We anticipate that by including the new terms, or by studying the full potential numerically as we will do in \S~\ref{ssec:nummetastates}, one can see that 
the metastable minimum will disappear
above a critical value of the anti-M2 charge. Above that value the
potential shows a perturbative instability toward one of the globally
supersymmetric minima described in \S~\ref{sec:susymin}, which are located at the right
boundary of the strip. 

Metastable probe M5 branes with M2 charge below the critical value can only decay 
non-perturbatively via tunneling to the globally supersymmetric minima. We postpone the discussion of this decay process to \S~\ref{ssec:decay} after discussing numerical results
regarding the vacuum structure of metastable probes in a simple LLM background in \S~\ref{ssec:nummetastates}.

The discussion regarding local minima in white strips close to even boundaries $x^{(2i)}$ is
completely analogous but, as discussed in \S~\ref{ssec:degsusy}, the role 
of M2 and anti-M2 branes are exchanged so that at even boundaries
anti-M2 branes are BPS and the supersymmetry breaking polarized M5 brane contains positive M2 brane charge.
One finds the same structure of metastable minima as before but now for small positive $p$.
To show this, one can start with the analogue
of~\eqref{approximationp} which is given by:
\begin{equation}
\mathcal{H}_+ \approx -p \left[ B_+(x) - \frac{1}{H_+(x)}\right] + c_3^+(x)\left[ B_+(x) - \frac{1}{H_+(x)}\right] -\frac{1}{\zeta_+^{2}(x)} + \frac{1}{p} \frac{1}{2\zeta_+^3(x)}
\, .
\end{equation}
Expanding in $2(x^{(i)}-x)\sim r^2$ one gets 
\begin{equation}\label{Hpgeneral2}
\mathcal{H_+} \approx -p \,a_1 + p \, \frac{a_2}{2} \, r^2 + \frac{1}{2p}r^6 \,.
\end{equation}
If $a_2<0$ the above expression minimizes at 
$r^2= p\,\sqrt{-\frac{a_2}{3}}$ and the discussion then proceeds as before.

\subsection{Numerical results}\label{ssec:nummetastates}

\begin{figure}[ht!]
\centering
\subfigure[$p=-1/2$.]{
 \includegraphics[width=0.43\textwidth]{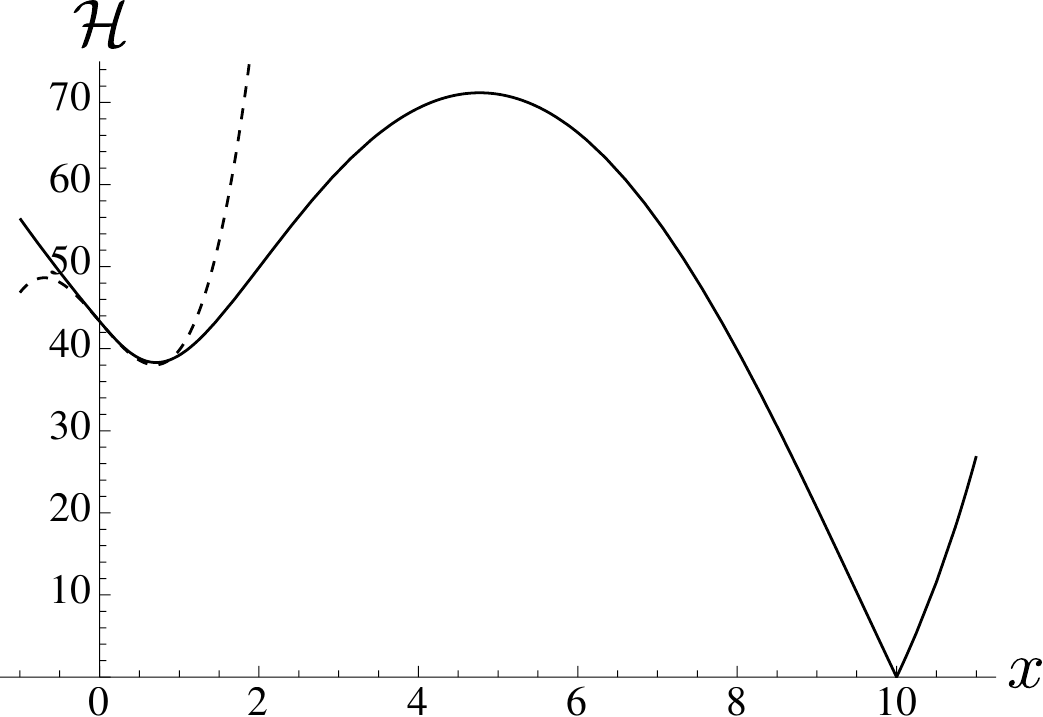}
 }
 \hspace{1cm}
 \subfigure[$p=-2$.]{
  \includegraphics[width=0.43\textwidth]{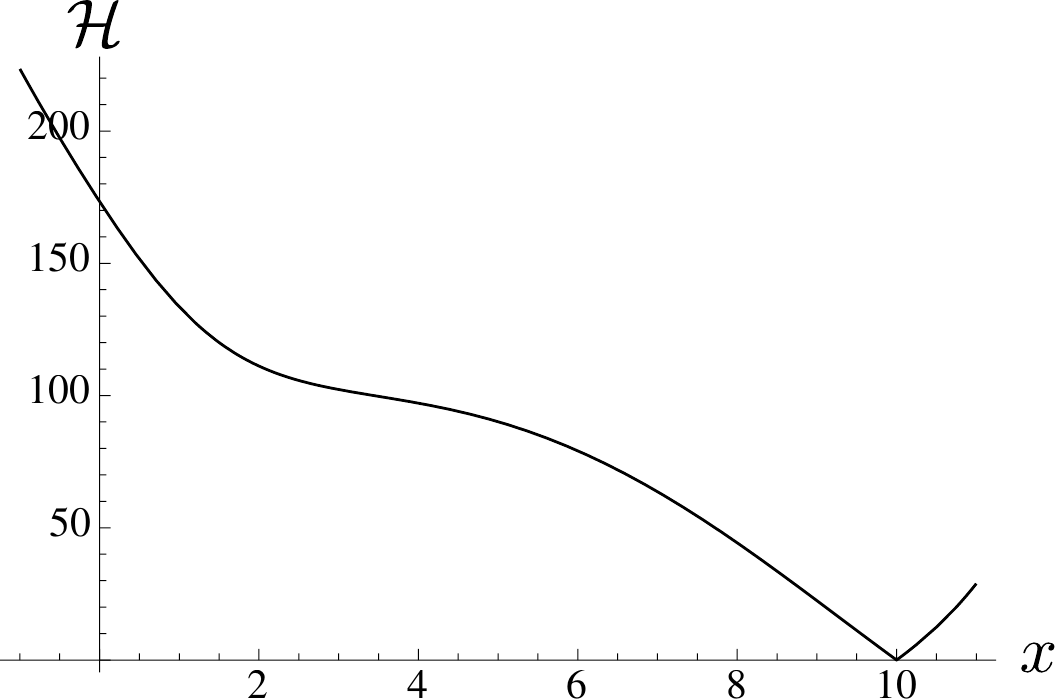}
 }
\caption{(a) Metastable minimum for negative $p$. The dashed line is the
  leading order approximation of the Hamiltonian as given in~\eqref{sec5HLO}. Below we give a Contour plot of \eqref{eq:Hamiltonian} in the $x-y$ plane which shows that the Hamiltonian indeed minimizes on the $y=0$ axis. (b) For larger
  $|p|$ the minimum disappears.}
\label{fig:metamin}
\end{figure}

\begin{figure}[ht!]
\centering
\subfigure[$p=-1/2$.]{
 \includegraphics[width=0.4\textwidth]{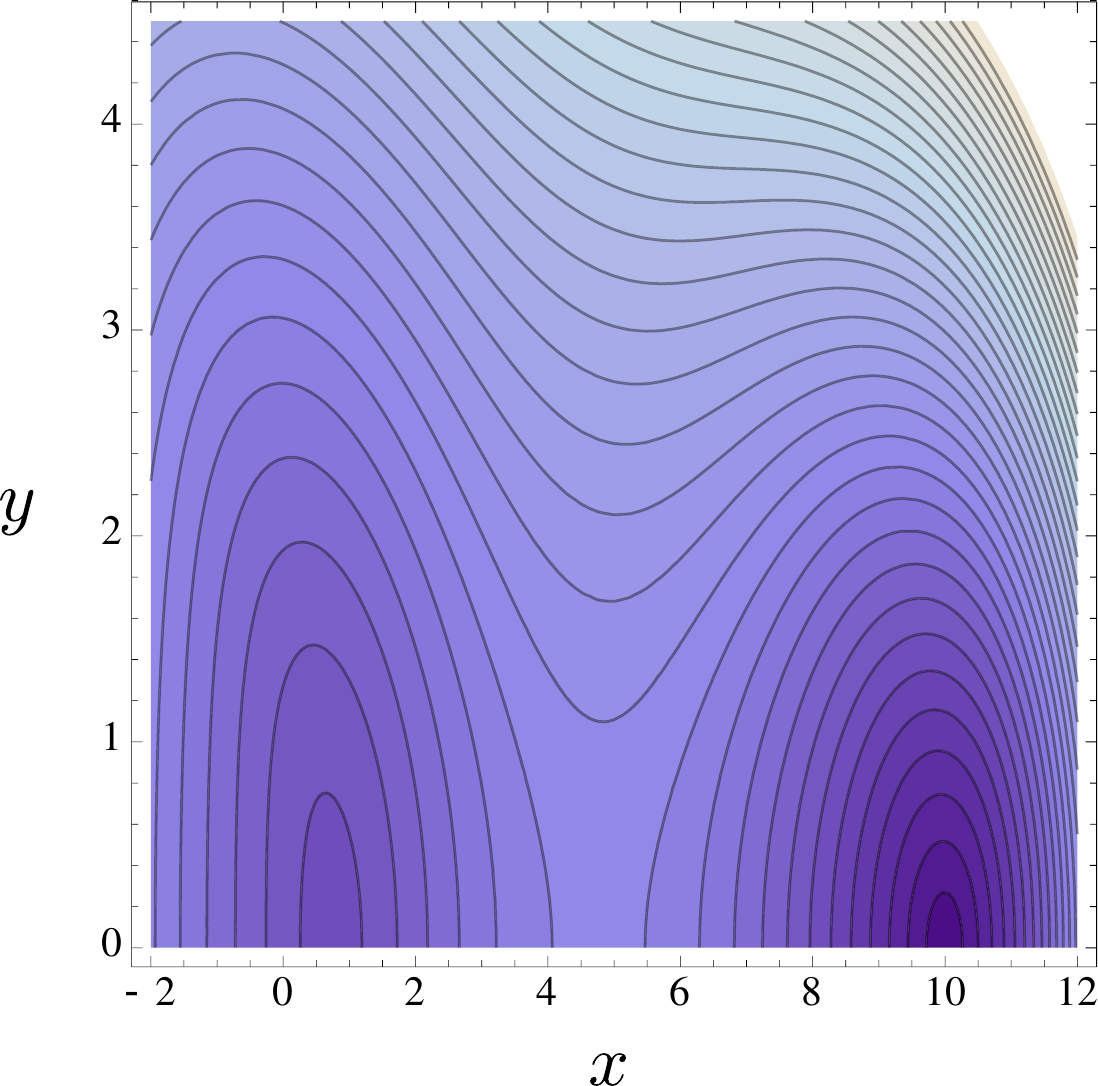}
 }
 \hspace{1.5cm}
 \subfigure[$p=-2$.]{
\includegraphics[width=0.4\textwidth]{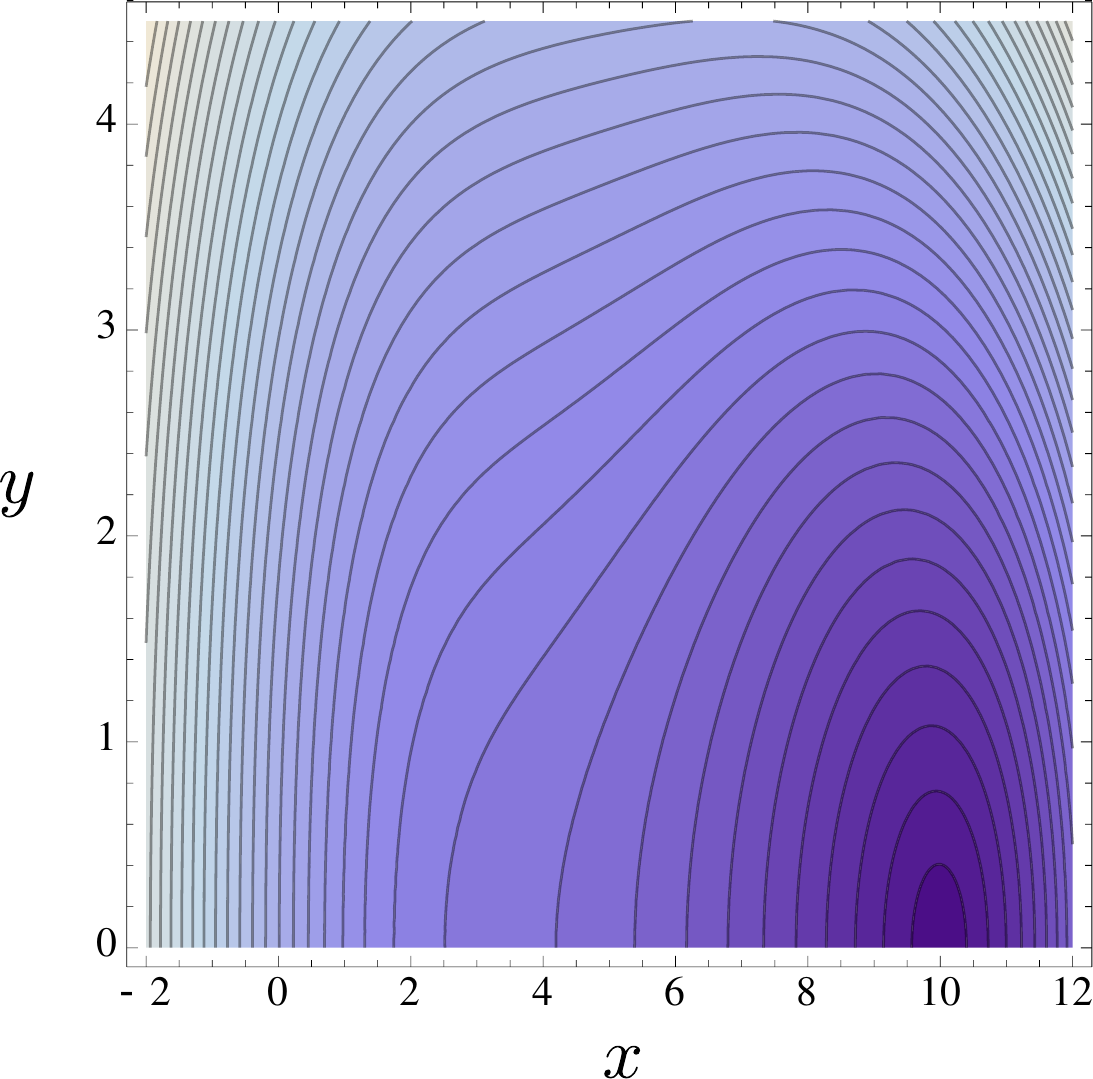}
 }
\caption{Contour plots in the $(x,y)$ plane of Figure \ref{fig:metamin}. 
Darker colors mean lower energy. (a) The metastable minimum (on the left) and the supersymmetric minimum (on the right) are at $y=0$. (b) The metastable minimum has disappeared and there is only the supersymmetric minimum (on the right) at $y=0$.}
\label{fig:contour}
\end{figure}

We now discuss the existence of metastable minima of the probe
Hamiltonian in the example of the single pair of white and black strips introduced in \S~\ref{sec:susymin}. We consider a probe M5 brane with induced anti-M2 charges
close to the left boundary of the finite-size white strip at $x=0$ (see Figure~\ref{fig:topology}) and we expand the Hamiltonian for small values of $x$.
The leading-order
approximation~\eqref{Hpgeneral} reduces to:
\begin{equation}\label{sec5HLO}
\mathcal{H}_+ \approx  |p| \,\frac{2 w(w+b)}{b}\,  -\,|p|\,\frac{3(2w+b)}{b} \,
x + \frac{4}{|p|} \,x^3
\, .
\end{equation}
It is easy to see that this potential has a metastable minimum at
\begin{equation}
 x_{meta} = \frac{|p|}{2} \sqrt{1+\frac{2 w}{b}}\, ,
\end{equation}
where the approximated potential~\eqref{sec5HLO} is
\begin{equation}
\mathcal{H}_+(x_{meta}) \approx \,|p|\,\frac{2w(w+b)}{b} - p^2\,
\left(1+\frac{2w}{b}\right)^{3/2} .
\end{equation}
We note that the terms in the potential~\eqref{sec5HLO} are detailed
balanced: at the minimum $x\sim |p|$ the last two terms scale
like $p^2$. This approximates well the potential for small $p$ and
small $x$, as shown in Figure~\ref{fig:metamin}(a). When $|p|$ increases,
the approximation breaks down and eventually the minimum disappears as shown in
Figure~\ref{fig:metamin}(b). 
We also plot in Figure~\ref{fig:contour} the full Hamiltonian~\eqref{eq:Hamiltonian} 
by keeping the dependence both on $x$ and $y$; one can easily see that
the Hamiltonian indeed minimizes at $y=0$.

To capture the transition from a metastable to an unstable
configuration at the critical value $p^\star$ of the anti-M2 charge, one could include higher order terms in
the expansion of the Hamiltonian. These are all the terms that, at the
minimum, scale with the same next-to-leading power of $p$. 
One can also study directly the zeroes of the derivative of the
potential numerically. We find that for the example $w=10$,
$b=3$, the transition happens around $p^{\star} \approx
-1.5$. We studied numerically the dependence of $p^{\star}$ on the
widths of the strips for various examples. One can easily show in this
way that increasing the width of the white strip in which the
metastable M5  brane polarizes, {\it i.e.} increasing the four-form flux $M_w$ on the
$S^4$, $|p^{\star}|$ grows and hence one can have a metastable M5 brane with
larger and larger number of anti-M2 branes dissolved in its
worldvolume. This is quite similar to~\cite{Kachru:2002gs,Klebanov:2010qs}.

We remark that even if $|p|>|p^{\star}|$, one can always find a
metastable probe minimum just by considering polarization into
multiple M5 branes, as discussed in \S~\ref{ssec:action}. In fact, one
can divide the $|p|$ anti-M2 branes in $m$ groups and make a single
group polarize. One obtains a configuration with $m$ M5 branes on top
of each other, polarized at a radius proportional to $|p|/m$. Hence, we
can achieve $|p|/m < |p^{\star}|$ by a suitable choice of $m$.

\subsection{Decay of metastable branes}\label{ssec:decay}

We have seen that for induced anti-M2 charge, the probe M5 brane has
locally stable minima at small but finite distance away from odd strip boundaries. 
These minima are classically stable since there is a non-perturbative barrier toward
the global supersymmetric minimum close to the other strip
boundary. Quantum mechanically, our probe will decay via bubble
nucleation to this supersymmetric minimum. We now briefly describe how
this process will take place. A similar mechanism was described
in~\cite{Kachru:2002gs,Klebanov:2010qs} but in the present case,
much like the supertube decays of~\cite{Bena:2011fc}, there is an
additional subtlety due to the presence of Dirac strings that we
would like to clarify. While we will present the decay process for
the example of the single pair of white and black strips it should
be understood that the discussion carries over to the decay of 
metastable probes placed in any strip of a general multi-strip configuration.

The decay of the metastable M5 brane probe can be understood as brane-flux annihilation 
of its induced anti-M2 charge against the M2 charge dissolved in the background flux.
Recall that the four-form flux through the four-sphere that stretches between the left and right boundary of the white strip and which contains the $S^3$ the M5 brane is wrapping is proportional to the size of the strip (see \S~\ref{ssec:M2M5charges}). We can write this as\footnote{Note that we drop all normalization factors in order to avoid cumbersome notation.}
\begin{equation}
M_{w} = \int_{x^{(1)}}^{x^{(2)} } d c_3^+= c_3^+({x^{(2)}}) - c_3^+({x^{(1)}})\,. 
\end{equation}
The M5 brane couples magnetically to $c_3^+$ and so, when it sweeps out the four-sphere $S^4$ from the North Pole to the South Pole, the amount of four-form flux through the orthogonal four-sphere $\tilde S^4$, given by $M_b$, changes by one unit.
Since we need at least two patches (the North Pole patch and the South Pole patch) to describe this process, we need to understand what happens to the probe when we change patch.

So far, we worked in a gauge where the three-form potential vanishes at the boundary of the strip that we are expanding around, which translates to fixing the constant $c$. This ensures that we work in a patch with no Dirac strings at that boundary and is thus the correct gauge in order to describe the physics of metastable minimum close to this boundary. 
When the metastable M5 brane tunnels to the stable minimum close to the other boundary, its quantized anti-M2 charge $p$ stays the same, but its effective anti-M2 charge 
\begin{equation}\label{peffdecay}
p^{eff}_+(x^{(i)})=p-c_3^+(x^{(i)})\,,
\end{equation}
changes. Without loss of generality we consider metastable probes close to the boundary $x^{(1)}$ of the white strip and gauge fix $c_3^+(x^{(1)})=0$. In this patch ``$1$'' we denote by $p_1\equiv p$ the quantized anti-M2 charge of the probe. The effective anti-M2 charge at the left boundary is  $p^{eff}_+(x^{(1)})=p$ while after the decay to the right boundary the effective anti-M2 charge is $p^{eff}(x^{(2)})=p-M_w$.
Once the probe M5 brane has tunneled to the supersymmetric minimum close to the boundary $x^{(2)}$ we need to change patch in order to correctly describe the physics at that minimum.
The gauge transformation parameter when changing from patch ``$1$'' (no Dirac strings at $x^{(1)}$) to patch ``$2$'' (no Dirac strings at $x^{(2)}$) is
\begin{equation}
 \gamma_{12}= c_3^+({x^{(1)}}) - c_3^+({x^{(2)}})=-M_{w} \,.
\end{equation}
When changing patch, the effective anti-M2 charge~\eqref{peffdecay} stays the same while the quantized anti-M2 charge changes according to
\begin{equation}
 p_2 = p_1+ \gamma_{12}=p-M_{w}\,,\label{ppatch}
\end{equation}
where $p_2$ denotes the quantized anti-M2 charge in the patch where there are no Dirac strings at the boundary $x^{(2)}$. Note that the change in the quantized anti-M2 charge after changing patch is the same as the change in the effective anti-M2 charge after the decay.

To summarize, in order to describe the vacuum structure and the dynamics of the probe one has to work in a fixed gauge and thus keep the quantized charges of the probe fixed. To describe the physics of the probe in a minimum close to the left/right boundary of a strip before and after the decay one has to work in a gauge where there are no Dirac strings at that boundary (North/South Pole of the four-sphere). 

In the decay process the quantized anti-M2 charge of the metastable probe changes according to \eqref{ppatch} by
\begin{equation}
 \Delta p = p_2-p_1 = -M_w\,.\label{Deltap}
\end{equation}
Furthermore, as anticipated above, when the probe sweeps out the
four-sphere between the boundaries $x^{(1)}$ and $x^{(2)}$ it changes
the four-form flux $M_b$ through the orthogonal four-sphere by one unit. Hence, the initial M2 charge dissolved in the background flux as given by $N_1= M_w M_b$ differs from the final M2 charge precisely by the amount \eqref{Deltap}. The final background M2 charge dissolved in flux is
\begin{equation}
N_2 = M_w ( M_b + 1) \, .
\end{equation}
Note that the number of anti-branes actually {\it increases} during the decay and so does the amount of background flux. It thus seems suitable to call this decay process brane-flux {\it creation}.
One can easily check that this decay process conserves the total M2 charge of the background as measured in the UV:
\begin{equation}
 N^{UV}=N^{IR} + N^{flux}\,,
\end{equation}
where $N^{IR}$ denotes the the M2 charge due to the presence of the probe brane and $N^{flux}$ denotes the M2 charge dissolved in the background fluxes. Before the decay $N^{UV}_1=p+ M_w M_b$ while after the decay $N^{UV}_2=p-M_w + M_w (M_b+1)=N^{UV}_1$.

When the metastable M5 brane probe close to the boundary $x^{(1)}$ decays to the degenerate supersymmetric minimum at the boundary $x^{(2)}$, the initial $|p|$ units of induced anti-M2 charge
become $|p-M_w|$ anti-M2 branes located at $x^{(2)}$. At this boundary anti-M2 branes are supersymmetric. According to the discussion of \S ~\ref{ssec:polsusy} the $|p-M_w|$ anti-M2 branes can polarize into a supersymmetric minimum inside the black strip adjacent to the boundary $x^{(2)}$. 
We can also consider the mirrored situation: probe M5 branes with small positive induced M2 charge $p$ which are metastable close to the boundary $x^{(2)}$ and decay to the degenerate supersymmetric minimum at the boundary $x^{(1)}$. The $p+M_w$ M2-branes are supersymmetric at this boundary and can further polarize into a supersymmetric minimum inside the semi-infinite black strip.

While so far we have discussed polarization of multiple (anti-) M2 branes into a single M5 brane we can also consider polarization into multiple M5 branes both for the initial metastable as well as the final supersymmetric configuration. Polarizing $|p|$ anti-M2 into $m$ metastable M5 branes wrapping the $S^3$ modifies the quantized anti-M2 charge after the decay to $p_2=p- m M_w$. Likewise, the flux through the orthogonal sphere changes, not by one, but by $m$ units so that the final M2 charged dissolved in the background flux is $N_2=M_w(M_b+m)$.
After the decay the $|p-m M_w|$ anti-M2 branes can further polarize into a single or multiple M5 branes. As discussed in \S~\ref{ssec:polsusy} polarization into multiple M5 branes wrapping the $\tilde S^3$ shifts the location of the supersymmetric minimum~\eqref{susy_minimum_exact}; hence one should always be able to find a supersymmetric minimum inside the black strip by considering polarization into multiple M5 branes.
Hence metastable M5 branes, after decaying in the $S^3$ channel to a degenerate minimum, can polarize into a smooth supersymmetric minimum in the $\tilde S^3$ channel.
The decay process thus corresponds to the tunneling of metastable M5 branes carrying (anti-) M2 charge to a supersymmetric minimum dual to a classical supersymmetric vacuum of the mass-deformed M2 brane theory.

\section{Discussion}\label{sec:discussion}

In this paper we probed bubbling AdS solutions holographically dual to the mass-deformed M2
brane theory. We studied the dynamics of probe M5 branes with dissolved
M2/anti-M2 branes, wrapping contractible three-cycles inside various
four-spheres in the background geometries.
For M5 branes with M2 brane charge parallel to the background flux we found 
supersymmetric global minima of the probe potential, which  explicitly demonstrate 
that the background geometries are indeed sourced by M5 branes shells with 
dissolved M2 charge. Moreover, we stress that the potential we derived in a 
fully backreacted M5 brane background is in agreement with the one obtained in~\cite{Bena:2000zb} from an analysis \`a la Polchinski-Strassler.

For M5 branes with $|p|$ units of M2 brane charge opposite to the background flux, we found
metastable configurations for small $|p|$ near left boundaries of
white strips of an LLM solution. Above a critical value, the
metastable minimum disappears and the M5 brane becomes unstable toward
perturbative decay to a supersymmetric state. This situation is
very similar to metastable probes in
Klebanov-Strassler~\cite{Kachru:2002gs}, CGLP backgrounds~\cite{Klebanov:2010qs}
and bubbling black hole microstate geometries~\cite{Bena:2011fc}. 
Since the BW and LLM geometries are dual to states of the mass-deformed M2 brane
theory, presumably described by a mass deformation of the ABJM
theory~\cite{Aharony:2008ug,Gomis:2008vc}, the solution corresponding
to our metastable probe M5 branes should be dual to a metastable state
in this theory. It would be clearly very interesting to understand this
better from the field theory side. 

By T-duality, our probes correspond to metastable giant gravitons in
the type IIB frame, namely D3 branes with angular momentum wrapping
one of the spheres of the LLM geometries. It would be interesting to
generalize our investigation to the full type IIB solution, described
by a generic configuration of black and white droplets on a plane.
We expect metastable configurations to exist in this case too.

We could also speculate that a similar result will hold in the yet to be
found gravity solution corresponding to the polarization of D3 branes into
D5 and NS5 branes in
\mbox{$AdS_5 \times S^5$}, which was studied in~\cite{Polchinski:2000uf}. This would point toward the existence of metastable
states in the $\mathcal{N}=1^{\star}$ SYM theory in four dimension, which is
obtained by giving masses to the three chiral multiplets of
$\mathcal{N}=4$ SYM theory. 

Finally, we believe that the most important open problem is to find the backreacted
solution corresponding to the metastable M5 branes. Since the
backgrounds we are probing correspond themselves to the backreaction
of M5 branes with M2 charge dissolved in flux, we believe that it should be possible to
extend some of the techniques recently used to study anti-branes
backreaction in flux compactifcations (see for
example~\cite{Borokhov:2002fm,Bena:2011wh,Bena:2012vz}) in order to construct the metastable
M5 gravity solution. 
The fully backreacted solution would be needed in order to check the local stability in 
the supergravity regime.
We note that in the probe approximation we detect a negative $r^2$
term in the polarization potential. In a fully backreacted regime, such a term would imply that the throat created by the anti-branes repels a fellow
probe anti-brane, thus signaling a tachyonic direction. Such an instability was found in~\cite{Bena:2014bxa} for anti-M2 branes in the CGLP background~\cite{Cvetic:2000db}.
If, in our case, the negative $r^2$ term persists in the backreacted regime, this would imply richer dynamics than the non-perturbative bubble nucleation picture indicated by the probe analysis.

Furthermore, since our result is quite similar to the
metastable supertubes found in~\cite{Bena:2011fc,Bena:2012zi} in the
probe approximation, computing the backreaction of our metastable M5 branes with dissolved M2 
charge could give insight into the more challenging non-BPS supertube
backreaction, and thus into the construction of large classes of 
non-extremal black hole microstate geometries in the context of 
the fuzzball proposal. The study of the stability of such
fully backreacted non-supersymmetric solutions would be relevant for understanding
the emission process from microstates and to compare with the semi-classical expectation.
We hope to come back to these problems in the near future.

\acknowledgments{We are grateful to Iosif Bena for useful discussions and comments on the manuscript.
  S.M. would like to thank the Isaac Newton Institute in Cambridge and 
  the organizers of the workshop ``Supersymmetry Breaking in String Theory''
  for hospitality while part of this work was completed. A.P. would like to thank the Aspen Center for Physics for hospitality in the final stage of this work.  
  The work of S.M. is supported by the ERC Advanced Grant 32004 -- Strings and Gravity. 
  The work of A.P. is supported by the National Science Foundation Grant No. PHY12-05500. The work of G.P. is supported in part by the ERC Starting Grant 240210, String-QCD-BH and by the Templeton Grant 48222: ``String Theory and the Anthropic Universe''.}

\appendix

\section{Review of type II bubbling geometries}\label{appsec:bubblinggeo}

In this section we review the type IIB geometries constructed by
Lin, Lunin and Maldacena (LLM) in \cite{Lin:2004nb}. We also perform a T-duality to obtain the
corresponding IIA solution and we compute the RR flux gauge
potentials explicitly. The uplift of the IIA solution to M-theory permits to
obtain the family of solutions which correspond to dielectric M2
brane vacua of the mass-deformed M2
theory~\cite{Bena:2000zb,Pope:2003jp,Bena:2004jw} presented in \S~\ref{sec:M}.

\subsection{Type IIB  solutions}\label{appssec:IIB}

The LLM type IIB solutions~\cite{Lin:2004nb} correspond to states of $\mathcal{N}=4$ SYM theory on $R\times S^3$. They preserve 16 supercharges and have an $SO(4)\times SO(4)\times R$ bosonic
symmetry, hence they contain two three-spheres $S^3$, $\tilde S^3$ and
a Killing vector. 
The metric and five-form flux compatible with such symmetries are:\footnote{The LLM function $H$
  in~\eqref{IIBAnsatz} should not be confused with the warp factor $H$
in~\eqref{11metric}.}
\begin{align}
ds^2 &= g_{\mu \nu} dx^\mu dx^\nu + e^{H+G} d\Omega_3^2 + e^{H-G} d\tilde \Omega_3^2\,, \label{IIBAnsatz} \\
F_5 &= F_{\mu \nu} dx^\mu \wedge dx^\nu \wedge d\Omega_3^2 + \tilde F_{\mu \nu} dx^\mu \wedge dx^\nu \wedge d\tilde \Omega_3^2\,,
\end{align}
where $\mu,\nu=0,...,3$ and $d\Omega_3^2$, $d\tilde \Omega_3^2$ denote the
metric on the three-spheres. The dilaton and axion are assumed to be
constant and the three-form field strengths are set to zero. 
Requiring that the above Ansatz preserves the Killing spinor equations
yields the following solution for the metric:
\begin{align}
ds^2  &= - h^{-2}(dt + V_i dx^i)^2 + h^2 (dy^2 + dx^idx^i) + y e^{G }
d\Omega_3^2 + y e^{ - G} d \tilde \Omega_3^2 \, ,
\end{align}
where $i=1,2$ and the functions $h,G,V$ are determined by a single
function $z$:
\begin{align}
h^{-2} &= 2 y \cosh G \, , \quad
G = {\rm arctanh}(2z)\,,\label{Gtozapp}\\
y \partial_y V_i &=\epsilon_{ij} \partial_j z\, , \quad
y (\partial_i V_j-\partial_j V_i) = \epsilon_{ij} \partial_y z\,. \label{eq:V2}
\end{align}
The five form flux is given by the two forms $F$, $\tilde F$ as
follows:
\begin{align}
F &= dB_t \wedge (dt + V) + B_t dV + d \hat B \,, \nn \\ \tilde F &=
d\tilde B_t \wedge (dt + V) + \tilde B_t dV + d \hat { \tilde B}\, ,
\end{align}
where we defined
\begin{align}
 B_t &= - \frac{1}{4} y^2 e^{2 G }\, , \quad  d \hat B =  -\frac{1}{4}
  y^3 \star_3 d A\, , \quad A=\frac{ z + \frac{1}{2}}{ y^2 }\,, \label{BthatBA}\\
\tilde B_t &= -  \frac{1}{4} y^2 e^{- 2 G}\, , \quad\;
d \hat {\tilde B} = -  \frac{1}{4} y^3 \star_3 d \tilde A\, , \quad
\tilde A= \frac{ z - \frac{1}{2}}{y^2 } \, ,\label{tildeBthatBA}
\end{align}
and the Hodge star $\star_3$ is referred to the flat space spanned by $y,x_1,x_2$.

The full solution is determined in terms of a single master function $z$ that obeys a linear equation:
\begin{equation}
 \partial_i \partial_i z + y \partial_y \left( \frac{\partial_y z}{y}
 \right) =0\, . \label{eq:zapp}
\end{equation}
The geometry described by this background is similar to that discussed in \S~\ref{ssec:Mtheoryuplift}: $y$ is the product of the radii of the three-spheres $S^3$ and $\tilde{S}^3$. The geometry is smooth if $z=\pm \frac{1}{2}$ on the $y=0$ plane spanned by $x_1$ and $x_2$. On this plane $S^3$ and $\tilde{S}^3$ shrink to a point in $z=-1/2$ and $z=1/2$ regions respectively, while both of them shrink on the boundaries of these regions. To represent a general solution one just needs to specify the black and white regions on the $y=0$ plane identified with the values $z=\pm 1/2$: see Figure~\ref{fig:blackwhite} (a) for an example.

\subsection{Type IIA solutions}\label{appssec:IIA}

We now T-dualize the IIB background~\eqref{IIBAnsatz}  along $x_1$. We assume that $V_2=0$ and that $V_1$ and $z$ do not depend on $x_1$. In the following we will drop the indices of $V_1$ and $x_2$ for convenience and rename $x_1=\omega_1$.
In the IIA frame the metric and the fluxes become\footnote{Note that the solution for the four-form field strength (D.1) as given in~\cite{Lin:2004nb} is incorrect. Consequently, also the solution for the four-form flux $G_4$ of the gravity dual of the mass-deformed M2 brane theory as stated in (2.35) of~\cite{Lin:2004nb} is incorrect. The correct form of $G_4$ is given in~\eqref{11G4}. In both~\eqref{11G4} and~\eqref{eq:F4} we dropped a factor $1/4$ due to different conventions for the volume forms on the spheres with respect to~\cite{Lin:2004nb}.}
\begin{align}
ds^2_{IIA}&= H^{-1} (-dt^2 + d\omega_1^2) + h^2(dy^2 + dx^2) +  y e^{G
} d\Omega_3^2 + y e^{ - G} d \tilde \Omega_3^2\, ,\label{IIAmetric}\\
B_2 &= -H^{-1} h^{-2} V dt \wedge d\omega_1\, ,\\
F_4 &=  \left[d(y^2 e^{2G} V) - y^3 \star_2 dA\right]
\wedge d\Omega_3 +  \left[d(y^2 e^{-2G} V) - y^3 \star_2
  d\tilde A\right] \wedge d\tilde \Omega_3\label{eq:F4} \, ,
\end{align}
where we defined the warp factor $H$ as:
\begin{equation}
H=e^{-2\Phi} = h^2-V^2 h^{-2}\, .
\end{equation}
The six-form field strength $F_6$ is given by $F_6 = \star F_4$.\footnote{We use conventions in which $\star F_4 =  F_6 =  dC_5 + H_3 \wedge C_3$.}
We obtain:
\begin{align}
 \star F_4 &=  H^{-1}  e^{3G} dt \wedge d\omega_1 \wedge
 \Big[ \star_2 d(y^2 e^{-2G} V) + y^3 d\tilde A \Big] \wedge
 d\Omega_3 \nn \\
 & \quad - H^{-1}  e^{-3G} dt \wedge d\omega_1 \wedge \Big[\star_2
   d(y^2 e^{2G} V) + y^3 dA \Big]\wedge d\tilde \Omega_3 \, .
\end{align}
For the computation of the polarization potential in \S~\ref{sec:polarization} we need the explicit expressions for
the RR gauge potentials $C_3$ and $C_5$. We define
 \begin{align}
C_{3} &=   c_3(x,y)d\Omega_3 + \tilde c_3(x,y) d\tilde \Omega_3\, ,\label{IIAC3} \\
C_{5} &= dt \wedge d\omega_1 \wedge \Big[ c_5(x,y) d\Omega_3 + \tilde
c_5(x,y) d\tilde \Omega_3\Big] \, .\label{IIAC5}
\end{align}
Since $C_1=0$ we have $F_4=dC_3$. It is useful to define
$\gamma_3=c_3-x - y^2 e^{2G} V+c$, where $c$ is an integration
constant that corresponds to the gauge choice for the three-form potential. The equation for $C_3$ along the $S^3$ becomes
\begin{equation} \label{eq:barc3}
d\gamma_3= -  \left(y^3 \star_2 dA + dx\right)\, ,
\end{equation}
which in components gives:
\begin{align}
\partial_y \gamma_3 &=y\partial_x z \nn\\
\partial_x \gamma_3 &=2z-y\partial_y z \, .\label{fundamentalsystem}
\end{align}
Note that to obtain $C_3$ we only have to solve this linear system.
With the explicit form for $z$ and $V$ in the
multi-strips solution \eqref{multiz}-\eqref{multiV} it is easy to
find an analytic solution, whose general form 
\begin{equation}\label{gamma3}
\gamma_3=  \sum_{i=1}^{2n+1} (-1)^{i+1} \gamma_3^0(x-x^{(i)},y)\,,
\end{equation}
is obtained by superpositions of the plane wave solution:
\begin{equation}
\gamma_3^0 = \frac{2 x^2+y^2}{2\sqrt{x^2+y^2}}\, .
\end{equation}
In an analogous way one obtains $C_3$ along $\tilde{S}^3$: to integrate $\tilde{c}_3$ one defines $\tilde{\gamma}_3=\tilde{c}_3+x-y^2 e^{-2G}V +\tilde{c}$ where $\tilde{\gamma_3}$ satisfies a linear system of equations identical to~\eqref{eq:barc3} and hence, up to integration constants, we get $\tilde{\gamma}_3=\gamma_3$. 

The equations for $C_5$ are obtained from the gauge-invariant improved field strength
$ F_6 =  dC_5 + H_3 \wedge C_3$. Defining $ \gamma_5=  c_5-c_3 H^{-1} h^{-2} V$ the equation for the part of $C_5$ along $S^3$ becomes 
\begin{equation}\label{eqgamma5}
 d\gamma_5 =  H^{-1} \left[- h^{-2} V \left(d(y^2 e^{2G} V) -y^3 \star_2 dA\right) + e^{3G} \left( \star_2 d(y^2 e^{-2G} V) + y^3d\tilde A \right)\right]\,,
\end{equation}
 which, remarkably, can be solved in closed form:
\begin{equation}\label{gamma5}
 \gamma_5 = \frac{2y^2}{1-2 z(x,y)}- y^2\,.
\end{equation}
In an analogous way one obtains $C_5$ along $\tilde S^3$: to integrate $\tilde c_5$ one defines $\tilde{\gamma}_5=\tilde{c}_5-\tilde{c}_3 H^{-1}h^{-2}V$ where $\tilde \gamma_5$ satisfies an equation identical to \eqref{eqgamma5} if one exchanges $G\leftrightarrow -G$ and $A \leftrightarrow \tilde A$ and the solution is given by~\eqref{gamma5} if one replaces $z \to -z$.

\section{Solution in the limit $y \to 0$}\label{appsec:limits} 

In the following we report the formulas for the $y\rightarrow0$ limit, keeping in mind that the background~\eqref{IIAmetric}-\eqref{eq:F4} is non-singular. 
While the limit has to be performed distinguishing between white and black strips, it can be shown that
 $V$ defined in~\eqref{eq:V2} and $\gamma_3$ defined in~\eqref{gamma3}  are well defined even for $y=0$, regardless of the particular strip considered.  
 The Hamiltonian (\ref{eq:Hamiltonian}) for the M5 brane probe is continuous for $y\rightarrow0$ even at the boundaries $x^{(i)}$ of the strips.
 
 \subsection*{White strips $z=1/2$}\label{appssec:limitwhite}

On white strips $S^3$ retains a finite-size, while $\tilde{S}^3$ shrinks to a point. Using equation~\eqref{Gtozapp}:
\begin{equation}
 z(x)=\frac{1}{2}\textmd{tanh} \, G(x) \, ,
\end{equation}
one obtains in the limit $y\to 0$ and $z\to +1/2$, using $e^G \to \infty$, the following expansion for the master function:\footnote{All the fields in the white strip limit will be marked with the subscript ``$+$'' .}
\begin{equation}
 z(x) \simeq 1/2 -e^{-2G(x)} \simeq 1/2 - y^2 \zeta_+^2(x) \, ,
\end{equation}
where $\zeta_+(x)$ is given by
\begin{equation}\label{equationc}
\zeta_+(x)=- \lim_{y\rightarrow 0} \frac{1}{\sqrt{2}}\frac{\partial_y
  z(x)}{\sqrt{1-2z(x)}} \, .
\end{equation}
For the multi-strip solutions~\eqref{multiz}-\eqref{multiV}, this function is given by
\begin{equation}
\zeta_+(x)=\frac{1}{2}\sqrt{\sum_{i=1}^{2s+1}(-1)^{i+1}\frac{|x-x^{(i)}|}{(x-x^{(i)})^3}}
\, .
\end{equation}
For the metric functions and the NS potential we get:
\begin{equation}
h_+(x) = \sqrt{\zeta_+(x)}\, ,\qquad H_+(x) = \zeta_+(x)-\frac{V_+^2(x)}{\zeta_+(x)}\, , \qquad B_+(x) =
-\frac{V_+(x)}{\zeta_+^2(x)-V_+^2(x)} \, ,
\end{equation}
where
\begin{equation}
 V_+(x) = \sum_{i=1}^{2s+1}  \frac{(-1)^{i}}{2 |x-x^{(i)}|}\,.
\end{equation}
The RR potentials on the finite $S^3$ become
\begin{equation}
c_3^{+}(x) = \frac{V_+(x)}{\zeta_+^2(x)}+\sum_{i=1}^{2s+1}(-1)^{i+1}|x-x^{(i)}|+x+c\,,\qquad c_5^{+}(x) = -c_3^{+}(x)B_+(x) +\frac{1}{\zeta_+^2(x)}\,.
\end{equation}
The RR potentials on the shrunk $\tilde{S}^3$ become
\begin{equation}
\tilde{c}_3^{+}(x) =\sum_{i=1}^{2s+1}(-1)^{i+1}|x-x^{(i)}|-x+\tilde{c}\,,\qquad \tilde{c}_5^{+}(x) = -\tilde{c}_3^{+}(x)B_+(x)\, .
\end{equation}
Note that $\tilde c_3^+(x)$ is constant inside a white strip.\\

We are interested in the potential for a probe brane wrapping either of the two three-spheres inside a white strip at $y=0$.
\begin{itemize}
\item The Hamiltonian inside a white strip for the probe M5 brane wrapping the finite-size $S^3$ is then given by:
\begin{equation}
 \mathcal{H}_+(x)=H_+^{-1}(x)\sqrt{\frac{H_+(x)}{\zeta_+^3(x)}+\left[ p-c_3^{+}(x)\right]^2}-p
 B_+(x)-c_5^{+}(x)\, .
\end{equation}
\item The Hamiltonian inside a white strip for the probe M5 brane wrapping the shrinking $\tilde S^3$ is given by
\begin{equation}\label{Hamiltonianslimit}
\tilde{\mathcal{H}}_+(x)=\frac{1}{\zeta_+^2(x)-V_+^2(x)}\left[ \zeta_+(x)
   |p-c_3^{+}(x)|+V_+(x) \left[ p-c_3^{+}(x)\right]
 \right] \, .
\end{equation}
\end{itemize}

 \subsection*{Black strips $z=-1/2$}\label{appssec:limitblack}

On black strips $S^3$ shrinks to a point, while $\tilde{S}^3$ retains a finite size. Proceeding as above one obtains in the limit $y\to 0$ and $z\to -1/2$, using $e^{-G} \to \infty$, the following expansion for the master function:\footnote{All the fields in the black strip limit will be marked with the subscript ``$-$'' .}
$$ z(x) \simeq -1/2 + e^{2G(x)} \simeq -1/2+y^2\zeta_-^2(x) $$
where $\zeta_-(x)$ is 
given by
\begin{equation}\label{equationd}
\zeta_-(x)=\lim_{y\rightarrow 0} \frac{1}{\sqrt{2}} \frac{\partial_y z(x)}{\sqrt{1+2z(x)}}
\end{equation}
For the multi-strip solutions~\eqref{multiz}-\eqref{multiV}, this function is given by
\begin{equation}\label{app:zetaminus}
\zeta_-(x)=\frac{1}{2}\sqrt{-\sum_{i=1}^{2s+1}(-1)^{i+1}\frac{|x-x^{(i)}|}{(x-x^{(i)})^3}}\,.
\end{equation}
We get for the metric functions and the NS potential 
\begin{equation}
h_-(x) = \sqrt{\zeta_-(x)}\,,\qquad H_-(x) = \zeta_-(x) - \frac{V^2_-(x)}{\zeta_-(x)}\,, \qquad B_-(x) = \frac{-V_-(x)}{\zeta_-^2(x)-V_-^2(x)}\,,
\end{equation}
where
\begin{equation}
 V_-(x) = \sum_{i=1}^{2s+1}  \frac{(-1)^{i}}{2 |x-x^{(i)}|}\,.
\end{equation}
The RR potentials on the finite $\tilde{S}^3$ become
\begin{equation}
\tilde{c}_3^{-}(x) = \frac{V_-(x)}{\zeta_-^2(x)}+\sum_{i=1}^{2s+1}(-1)^{i+1}|x-x^{(i)}| -x+\tilde{c}
\,,\qquad \tilde{c}_5^{-}(x) = -\tilde{c}_3^{-}(x) B_-(x) +\frac{1}{\zeta_-^2(x)} \,.
\end{equation}
The RR potentials on the shrinking $S^3$ become
\begin{equation}
c_3^{-}(x) = \sum_{i=1}^{2s+1}(-1)^{i+1}|x-x^{(i)}|+x+c\,,\qquad c_5^{-} = -c_3^{-}(x)B_-(x) \, .
\end{equation}
Note that $c_3^-(x)$ is constant inside a black strip.\\

We are interested in the potential for a probe brane wrapping either of the two three-spheres inside a black strip at $y=0$.
\begin{itemize}
 \item The Hamiltonian inside a black strip for the probe M5 brane wrapping the finite-size $\tilde{S}^3$ is then given by:
\begin{equation}
\tilde{\mathcal{H}}(x)_-=H_-^{-1}(x)\sqrt{\frac{H_-(x)}{\zeta_-^3(x)}\left[p-\tilde{c}_3^{-}(x)\right]^2}-p
 B_-(x)-\tilde{c}_5^{-}(x) \, .
\end{equation}
\item The Hamiltonian inside a black strip for the probe M5 brane wrapping the shrinking $S^3$ is given 
by
\begin{equation}\label{Hamiltonianslimit}
 \mathcal{H}_-(x)=\frac{1}{\zeta_-^2(x)-V_-^2(x)}\left[ \zeta_-(x)
   |p-c_3^{-}(x)|+V_-(x) \left[ p-c_3^{-}(x)\right]
 \right] \, .
\end{equation}
\end{itemize}

\section{Relation between LLM and BW}\label{app:BW}

In this section we provide a dictionary that relates the M-theory
solution of Lin, Lunin and Maldacena~\cite{Lin:2004nb} described in \S~\ref{ssec:Mtheoryuplift} to the solution of Bena and Warner~\cite{Bena:2004jw}.
The BW metric is written as
\begin{align}\label{metricBW}
ds^2_{11}&=16L^4e^{2B_0}(-dt^2+d\omega_1^2+d\omega_2^2)+e^{2B_2-B_0}(du^2+dv^2)  \nonumber \\ &\qquad +\frac{1}{4}e^{2B_3-B_0}u^2 \sigma_i \sigma_i
       +\frac{1}{4}e^{-2B_3-B_0}v^2 \tau_i \tau_i \,  ,
\end{align}
where $B_0$, $B_2$, $B_3$ are functions of $u$ and $v$ only and $\sigma_i$ and $\tau_i$ are left-infariant 1-forms that parameterize the two three-spheres.
Identifying
\begin{equation} \frac{1}{4} \sigma_i \sigma_i = d\Omega_3\, ,\qquad
  \frac{1}{4} \tau_i \tau_i = d\tilde{\Omega}_3 \, ,
\end{equation}
and comparing~\eqref{metricBW} with~\eqref{IIAmetric} one gets:
\begin{align}
 4u^2L^2 e^{2B_3} &= ye^G    \, , \label{eq1} \\
 4v^2L^2e^{-2B_3} &= ye^{-G}  \, , \nn\\
 4uvL^2 &=y \, , \nonumber \\
 2L^2(u^2-v^2) &= x\, , \nonumber \\
 \frac{1}{64L^6}e^{-3B_0} &= H \, .\nn
\end{align}
This yields the relation between the $(v,u)$ and $(y,x)$ coordinates of respectively BW and LLM: 
\begin{equation}\label{masterconversion}
  4L^2u^2=x+\sqrt{x^2+y^2}\, , \qquad 4L^2v^2= -x+\sqrt{x^2+y^2} \, .
\end{equation}
In BW the background fields are determined once one fixes a master
function $g(u,v)$ which satisfies the linear equation
\begin{equation}\label{eq:g}
\frac{\partial^2 g}{\partial u^2}+\frac{\partial^2 g}{\partial v^2}
-\frac{1}{u}\frac{\partial g}{\partial u}
-\frac{1}{v}\frac{\partial g}{\partial v} = 0 \, .
\end{equation} 
This is analogous to determining the master function $z(y,x)$ in the type IIA LLM background.
Using (\ref{masterconversion}) 
it is possible to rewrite $g(u,v)$ in terms of the $(y,x)$ coordinates. Considering that 
\begin{equation}
 z = \frac{1}{2}\frac{e^{2G}-1}{e^{2G}+1} \, ,
\end{equation}
and using (\ref{eq1}) one gets:
\begin{equation}
z=-2\partial_x g + z_0 \, ,
\end{equation}
where $z_0 =\frac{x}{2\sqrt{x^2+y^2}}$ is the half-filled plane
solution. With this identification one can check that equation~\eqref{eq:z} for $z$ is
equivalent to the master equation~\eqref{eq:g}. \\
The background of~\cite{Bena:2004jw} that preserves 16 supercharges also depends on the constant $\beta$. The latter is related to the mass deformation in the dual M2 brane theory. For $\beta \to 0$ the background reduces to the standard Coulomb branch of M2 branes only. The IIA background derived from~\cite{Lin:2004nb} has a fixed value $\beta=1/4$ and hence also the mass-deformation is fixed.

\bibliographystyle{utphys}

\providecommand{\href}[2]{#2}\begingroup\raggedright\endgroup

\end{document}